\documentclass[a4paper,12pt]{article}
\usepackage[left=20mm,right=20mm,top=20mm,bottom=30mm]{geometry}
\usepackage[affil-it]{authblk}
\usepackage{amsmath}
\usepackage{amssymb}
\usepackage{hyperref}
\usepackage{mathtools}
\usepackage{microtype}
\usepackage{tikz}
\usepackage{listings}
\usepackage{lmodern}
\usepackage[section]{placeins}
\usepackage{cite}

\usetikzlibrary{calc}
\usetikzlibrary{math}
\usetikzlibrary{arrows.meta}

\usepackage{tikz}

\usetikzlibrary{arrows}
\usetikzlibrary{decorations.pathmorphing}
\usetikzlibrary{decorations.markings}
\usetikzlibrary{calc}

\tikzset{
  >=stealth',
  midarrow/.style={
    postaction={
      decorate,
      decoration={markings, mark=at position .55 with {\arrow{>}}}
    }
  },
  fermion/.style=midarrow,
  photon/.style={
    decorate,
    decoration={snake, amplitude=2pt, segment length=8pt}
  },
  boson/.style={
    decorate,
    decoration={snake, amplitude=2pt, segment length=8pt}
  },
  gluon/.style={
    decorate,
    decoration={coil, amplitude=4pt, segment length=5pt}
  },
  scalar/.style=densely dashed,
  arrowsnake/.style={
    preaction={photon, draw},
    postaction=midarrow
  }
}

\hypersetup{
  colorlinks,
  linkcolor={blue!50!black},
  citecolor={blue!50!black},
  urlcolor={blue!50!black}
}

\lstdefinestyle{c++}{
  language=C++,
  basicstyle=\small \tt,
  keywordstyle=\bfseries,
  columns=fullflexible,
  numbers=left,
  showstringspaces=false
}

\lstdefinestyle{python}{
  language=Python,
  basicstyle=\small \tt,
  keywordstyle=\bfseries,
  columns=fullflexible,
  numbers=left,
  morekeywords={with},
  showstringspaces=false
}

\DeclareMathOperator{\asinh}{asinh}

\newcommand{\abs}[1]{\lvert #1 \rvert}
\newcommand{\nuc}[2]{$^{#2}$#1}

\begin{document}

\title{libepa --- a C++/Python library for calculations of cross sections of
ultraperipheral collisions}

\author[1]{E.~V.~Zhemchugov\footnote{\texttt{zhemchugovev@lebedev.ru}}}
\author[1]{S.~I.~Godunov}
\author[1]{E.~K.~Karkaryan}
\author[1]{V.~A.~Novikov}
\author[2]{A.~N.~Rozanov}
\author[1]{M.~I.~Vysotsky}

\affil[1]{
  \small I.~E.~Tamm Department of Theoretical Physics, Lebedev Physical
  Institute, \newline
  53 Leninskiy Prospekt, Moscow, 119991, Russia
}
\affil[2]{
  \small Centre de Physique de Particules de Marseille (CPPM), Aix-Marseille
  Universite, CNRS/IN2P3, \newline
  163 avenue de Luminy, case 902, Marseille, 13288, France
}

%\date{\today \\ (draft)}
\date{}

\maketitle

\begin{abstract}
The library provides a set of C++/Python functions for computing cross sections
of ultraperipheral collisions of high energy particles under the equivalent
photons approximation. Cross sections are represented through multiple integrals
over the phase space. The integrals are calculated through recurrent application
of algorithms for one dimensional integration. The paper contains an
introduction to the theory of ultraperipheral collisions, discusses the library
approach and provides a few examples of calculations.
\end{abstract}


{ \it PROGRAM SUMMARY }
\begin{itemize}
  \item \textit{Program title:} \texttt{libepa}.
  \item \textit{Developer's repository link:}
    \href{https://github.com/jini-zh/libepa}{https://github.com/jini-zh/libepa}.
  \item \textit{Licensing provisions:} GNU General Public License 3 (GPL3).
  \item \textit{Programming Language:} C++, Python.
  \item \textit{Supplementary material:} attached.
  \item \textit{Nature of the problem:} Ultraperipheral collisions of
  ultrarelativistic particles are a source of high energy photon-photon
  collisions.  Ultraperipheral collisions occur at particle accelerators, in
  particular the Large Hadron Collider. They can be used to discover new
  electrically charged particles or other particles that can be produced in
  photons fusion (e.g., axions), and to make precision tests of the Standard
  Model of particle physics. \texttt{libepa} is a tool to calculate cross
  sections of ultraperipheral collisions with restrictions on the phase space of
  the produced particles typical to high energy experiments.
  \item \textit{Solution method:} Cross sections are expressed in terms of
  multiple integrals over the phase space parameters and numerically calculated
  through recurrent application of algorithms for one-dimensional integration.
  Functional programming approach is used to simplify the interface and optimize
  the calculations.
\end{itemize}

\section{Introduction}

When a charged particle is accelerated to a speed comparable to the speed of
light, its electric field undergoes Lorentz contraction and looks like a thin
disc perpendicular to the direction of the particle motion. Under the equivalent
photons approximation (EPA, also known as the Weizs\"acker-Williams
approximation~\cite{zphys29-315, landau-epa, zphys88-612, williams-epa}), such
a field can be represented as a bunch of high energy photons distributed
according to a known spectrum. Consider a collision of two charged particles
passing at a distance from each other and colliding with their electromagnetic
fields. In this collision new particles can be produced through fusion of the
photons emitted from the fields of original particles. If both particles remain
intact after the collision (such collisions are called ultraperipheral (UPC)),
then the final state of the event will be just the two original particles and
whatever has been produced through photon fusion. This is in contrast to a
typical high energy hadron collision with dozens of particles in the final
state. 

Ultraperipheral collisions happen aplenty at any particle collider and they
allow for studying photon-photon collisions and physics of intense (though
short-lived) electromagnetic fields. For example, light by light scattering
($\gamma \gamma \to \gamma \gamma$) was recently observed~\cite{1702.01625,
1810.04602} and then its cross section was measured~\cite{2008.05355} by the
ATLAS and CMS collaborations at the Large Hadron Collider (LHC). The reaction
$\gamma \gamma \to W^+ W^-$ provides an opportunity to study a quartic coupling
of the Standard Model lagrangian and is also searched for at the
LHC~\cite{2010.04019, 2211.16320}. An interesting feature of ultraperipheral
collisions is that they can be used to search for any new hypothetical particle
coupling with the electromagnetic field. This is of particular interest to
supersymmetry which predicts many new electrically charged particles. Chargino
pair production reaction $\gamma \gamma \to \tilde \chi^+ \tilde \chi^-$ might
be useful to explore the compressed mass scenario in supersymmetric theories
(when the masses of the lightest chargino and neutralino are approximately
equal)~\cite{1906.08568}.

The usual approach to calculate cross sections of ultraperipheral collisions is
Monte Carlo~\cite{1102.2531, 1607.03838, 2007.12704, 2207.03012}.  However, in
the case when just a pair of a particle and its antiparticle is produced in an
ultraperipheral collision, the expression even for the fiducial cross section is
simple enough to be written explicitly in terms of a multiple integral over the
phase space. Often this integral can be calculated through common numerical
algorithms. The possibility to work with analytical expressions coupled with the
ability to quickly evaluate these expressions can provide valuable insights into
the phenomenology of the studied process. \texttt{libepa}, a library for
calculation of cross sections of ultraperipheral collisions of high energy
particles was created to help with the evaluation part of this
research~\cite{libepa}. 

This paper describes the physical model of ultraperipheral collisions used by
\texttt{libepa} in Section~\ref{s:physics}, its approach to setup and perform
calculations in Section~\ref{s:library}, discusses several examples of
phenomenological calculations that were used to provide theoretical description
of experiments at the LHC in Section~\ref{s:examples} and compares the library
to other programs solving the same problem in Section~\ref{s:validation}. This
is a further development of the papers~\cite{1806.07238, 2106.14842}.

\section{Physics}

\label{s:physics}

\subsection{UPC cross section}

\label{s:upc-xsection}

Electric field of a charged particle at rest can be represented as a bunch of
virtual photons with zero energy: $q' = (0, q'_x, q'_y, q'_z)$, where $q'$ is
the photon 4-momentum in the particle rest frame. In the reference frame where
the particle is moving along the $z$ axis with the velocity $v$, the photon
momentum is
\begin{gather}
  q = (\omega, \vec q_\perp, q_\parallel), \\
  \omega = \sqrt{\gamma^2 - 1} \, q'_z, \ 
  \vec q_\perp = (q'_x, q'_y), \ 
  q_\parallel = \gamma q'_z,
  \notag
\end{gather}
where $\gamma = 1 / \sqrt{1 - v^2}$ is the Lorentz factor of the
particle,\footnote{
  We work in the system of units such that $\hbar = c = 1$, where $\hbar$ is the
  Planck constant and $c$ is the speed of light.
}
$\omega$ is the photon energy, $\vec q_\perp$ is its transversal momentum,
$q_\parallel$ is its longitudinal momentum. When the particle velocity is close
to the speed of light, $\gamma \gg 1$ and the photon energy $\omega \approx
q_\parallel \gg \sqrt{-q^2}$. The photon thus becomes approximately real. The
contribution of the photon virtuality to observable values is a function of the
photon 4-momentum squared which is conventionally denoted as
\begin{equation}
  Q^2 \equiv -q^2 = q'^2_x + q'^2_y + q'^2_z.
\end{equation}
For $\gamma \gg 1$,
\begin{equation}
  Q^2 \approx q_\perp^2 + (\omega / \gamma)^2.
  \label{photon-virtuality}
\end{equation}

\begin{figure}[!tb]
  \centering
  \begin{tikzpicture}[baseline=(O)]
    \coordinate (O) at (0, 1);
    \draw (-2, -1) -- (0, -1);
    \draw (0, -1) -- (2, -1);
    \draw [photon]  (0, -1) -- node [right] {$\gamma^*$} (0, 1);
    \draw [->] ([xshift=-2mm]$(0, -1)!0.25!(0, 1)$)
               to node [midway, left] {$q$}
               ([xshift=-2mm]$(0, -1)!0.5!(0, 1)$);
    \draw (-2, 1) -- (0, 1);
    \draw (0, 1) -- +( 15:2);
    \draw (0, 1) -- +(  5:2);
    \draw (0, 1) -- +( -5:2);
    \draw (0, 1) -- +(-15:2);
    \draw [fill=gray] (0, 1) circle (0.25);
  \end{tikzpicture}
  \begin{tikzpicture}[baseline=(O)]
    \coordinate (O) at (0, 0);
    \draw [->] (-1, 0) -- node [above] {EPA} node [below] {$\gamma \gg 1$} (1, 0);
  \end{tikzpicture}
  \begin{tikzpicture}[baseline=(O)]
    \coordinate (O) at (0, 1);

    \draw (-2, 2) -- (0, 1);
    \draw [photon] (-2, 0) node [below left] {$\gamma$} -- (0, 1);
    \draw [->] ([yshift=-2mm]$(-2, 0)!0.25!(0, 1)$)
               to node [midway, below] {$q$}
               ([yshift=-2mm]$(-2, 0)!0.5!(0, 1)$);
    \draw (0, 1) -- +( 15:2);
    \draw (0, 1) -- +(  5:2);
    \draw (0, 1) -- +( -5:2);
    \draw (0, 1) -- +(-15:2);
    \draw [fill=gray] (0, 1) circle (0.25);
  \end{tikzpicture}
  \caption{
    \textit{Left:} Feynman diagram depicting a process mediated by a photon
    emitted by the electric field of an ultrarelativistic particle.
    \textit{Right:} Under the equivalent photon approximation, the virtual
    photon is replaced by a real photon.
  }
  \label{f:epa-virtual}
\end{figure}
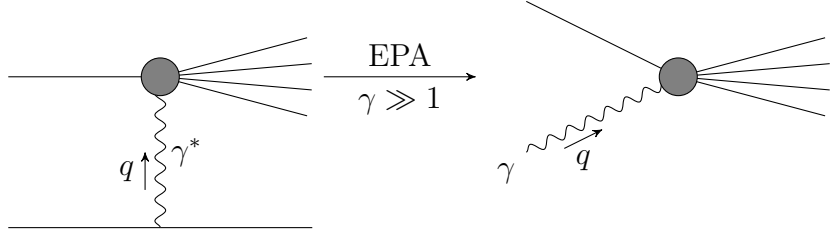

Electromagnetic interactions induced by the electric field of a relativistic
particle are described by Feynman diagrams each containing a photon propagating
between the particle line and the rest of the diagram, see
Fig.~\ref{f:epa-virtual}. Under the equivalent photon approximation (EPA), the
virtual photon can be replaced with a real photon with its momentum distributed
according to a known function $N(q)$ depending on the nature of the source
particle. Cross section for the original process $\sigma$ can then be calculated
through the cross section for the process with the real photon $\sigma_\gamma$
as follows:
\begin{equation}
  \sigma = \int \sigma_\gamma \, N(q) \, \mathrm{d}^3 q.
\end{equation}
Integration over $\mathrm{d}^2 q_\perp$ yields
\begin{equation}
  \sigma = \int \sigma_\gamma \, n(\omega) \, \mathrm{d} \omega,
\end{equation}
where
\begin{equation}
  n(\omega) = \int N(q) \, \mathrm{d}^2 q_\perp
  \label{epa-spectrum}
\end{equation}
is the equivalent photon spectrum for the source particle.

Consider a collision of two relativistic charged particles $A$ and $B$.
For now, let us assume that the particles only interact through the
electromagnetic field.  Let $n_A(\omega)$ and $n_B(\omega)$ be the equivalent
photon spectra of $A$ and $B$. Then the cross section for this collision is
\begin{equation}
  \sigma(AB \to AB X)
  = \int\limits_0^\infty \mathrm{d} \omega_1
    \int\limits_0^\infty \mathrm{d} \omega_2
    \, \sigma(\gamma \gamma \to X)
    \, n_A(\omega_1)
    \, n_B(\omega_2),
  \label{epa-xsection}
\end{equation}
where $X$ is the system produced in photon-photon fusion. 

In experimental particle physics it is common to apply constraints on the phase
space of the produced system, to suppress the background or as a feature of the
experimental setup. In the case of the production of a pair of particles, i.e.
when $X = \chi^+ \chi^-$, the two most common constraints are the cuts on the
transversal momentum $p_T$ and pseudorapidity $\eta$ of the produced
particles:
\begin{equation}
  p_T > \hat p_T, \ -\hat \eta < \eta < \hat \eta
  \qquad \text{(for each particle),}
  \label{constraints}
\end{equation}
where $\hat p_T$ and $\hat \eta$ are the cuts values. The constraint on the
transversal momentum is commonly used to suppress the background from less
energetic processes, while the constraint on the pseudorapidity is used to
select events with the particles hitting the sensitive regions of the detector.

There are also constraints specific to ultraperipheral collisions. In these
collisions the colliding particles survive and can be detected after the
collision. This is the case of proton-proton collisions in the ATLAS and CMS
detectors at the LHC: depending on their energy loss, the protons may hit
special detectors located a few hundred meters down the beam. The condition for
such events can be translated to constraints on the photons energies in the UPC: 
\begin{equation}
  \hat \omega_{1,\text{min}} < \omega_1 < \hat \omega_{1,\text{max}}, \ 
  \hat \omega_{2,\text{min}} < \omega_2 < \hat \omega_{2,\text{max}}.
  \label{constraints-upc}
\end{equation}

To write down the expression for the fiducial cross section, we change the
integration variables in~\eqref{epa-xsection} from the photons energies
$\omega_1$, $\omega_2$ to the invariant mass of the produced system $\sqrt{s} =
\sqrt{4 \omega_1 \omega_2}$ ($s$ is the Mandelstam variable for the reaction
$\gamma \gamma \to X$) and its rapidity $y = \tfrac12 \ln
\tfrac{\omega_1}{\omega_2}$.  In this notation the expression for the
differential total (not fiducial) cross section is
\begin{equation}
  \frac{\mathrm{d} \sigma(AB \to ABX)}{\mathrm{d} \sqrt{s}}
  = \sigma(\gamma \gamma \to X)
  \cdot \frac{\mathrm{d} L_{AB}}{\mathrm{d} \sqrt{s}},
  \label{xsection}
\end{equation}
where $L_{AB}$ is the photon-photon luminosity in the collision of particles $A$
and $B$,
\begin{equation}
  \frac{\mathrm{d} L_{AB}}{\mathrm{d} \sqrt{s}}
  = \frac{\sqrt{s}}{2}
    \int\limits_{-\infty}^\infty
         n_A \left( \frac{\sqrt{s}}{2} \mathrm{e}^y \right)
      \, n_B \left( \frac{\sqrt{s}}{2} \mathrm{e}^{-y} \right)
      \, \mathrm{d} y.
  \label{luminosity}
\end{equation}
Then the cut on $\eta$ can be taken into account by changing the limits of the
integration with respect to $y$. To establish the connection between the
rapidity of the system and pseudorapidities of the produced particles, we solve
the system of equations describing the conservation of energy and momentum in
the collision depicted at Fig.~\ref{f:collision}:
\begin{figure}[!tb]
  \centering
  \begin{tikzpicture}[>=stealth]
    \filldraw (0, 0) circle [radius=0.1];
    \draw [midarrow] (0, -2) node [below] {$\gamma$} -- (0, 0);
    \draw [midarrow] (0,  2) node [above] {$\gamma$} -- (0, 0);
    \draw [->] (0, 0) -- ( 10:2) node [right] {$\chi^+$};
    \draw [->] (0, 0) -- (-1.96, -1.29) node [below left] {$\chi^-$};
    \draw (0, 0.6) arc (90:50:0.6) node [above right] {$\theta_1$} arc (50:10:0.6);
    \draw (0, 0.8) arc (90:213:0.8);
    \draw (0, 0.85)
      arc (90:165:0.85) node [above left] {$\theta_2$}
      arc (165:213:0.85);
    \node [left] at (0, 1.5) {$\omega_2$};
    \node [right] at (0, -1) {$\omega_1$};
  \end{tikzpicture}
  \caption{Momenta of the particles participating in the $\gamma \gamma \to
  \chi^+ \chi^-$ reaction.}
  \label{f:collision}
\end{figure}
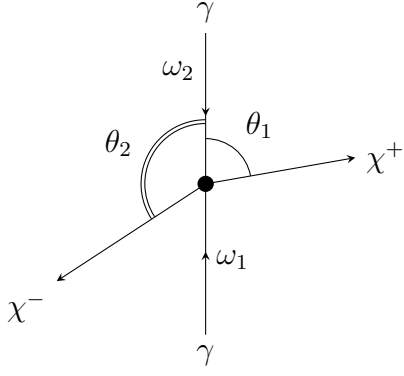
\begin{equation}
  \left\{
    \begin{aligned}
      \omega_1 + \omega_2
      &= \sqrt{\frac{p_T^2}{\sin^2 \theta_1} + m_\chi^2}
       + \sqrt{\frac{p_T^2}{\sin^2 \theta_2} + m_\chi^2},
      \\
      \omega_1 - \omega_2
      &= \frac{p_T}{\tan \theta_1} + \frac{p_T}{\tan \theta_2},
    \end{aligned}
  \right.
\end{equation}
where $m_\chi$ is the mass of $\chi^\pm$.  Recall that $\eta = -\ln \tan
(\theta/2)$ where $\theta$ is the polar angle of the particle momentum (when the
axis is along the beam), and $\omega_1 = \tfrac{\sqrt{s}}{2} \mathrm{e}^y$,
$\omega_2 = \tfrac{\sqrt{s}}{2} \mathrm{e}^{-y}$. With these substitutions, the
solution is
\begin{equation}
  \sinh \eta_{1,2}
  = \frac{\sqrt{s}}{2 p_T}
    \left(
      \sinh y \pm \cosh y \, \sqrt{1 - \frac{p_T^2 + m_\chi^2}{s/4}}
    \right).
  \label{eta-y}
\end{equation}
The constraint $-\hat \eta < \eta < \hat \eta$ then translates to
\begin{equation}
  y_+(-\hat \eta) < y < y_-(\hat \eta), \label{y-bounds-asymmetrical}
\end{equation}
where
\begin{equation}
  \begin{split}
    y_\pm(\eta)
    &= \asinh \left[
         \frac{p_T \sqrt{s} / 2}{p_T^2 + m_\chi^2}
         \left(
           \sinh \eta
           \pm   \sqrt{\cosh^2 \eta + \frac{m_\chi^2}{p_T^2}}
           \cdot \sqrt{1 - \frac{p_T^2 + m_\chi^2}{s / 4}}
         \right)
       \right]
    \\
    &= \ln \left(
         \frac{2 p_T}{\sqrt{s}}
         \cdot
         \frac{\sinh \eta + \sqrt{\cosh^2 \eta + \frac{m_\chi^2}{p_T^2}}}
              {1 \mp \sqrt{1 - \frac{p_T^2 + m_\chi^2}{s / 4}}}
       \right).
  \end{split}
\end{equation}
Since $y_+(-\eta) = -y_-(\eta)$, in the case of symmetrical bounds on
pseudorapidity considered here inequality~\eqref{y-bounds-asymmetrical}
simplifies to
\begin{equation}
  -\hat y < y < \hat y, \ \hat y \equiv y_-(\hat \eta).
\end{equation}

Note that $\hat y$ depends on $p_T$. To take this constraint into account, as
well as the constraint $p_T > \hat p_T$, we introduce into~\eqref{xsection} an
additional integration with respect to $p_T$:
\begin{gather}
  \frac{\mathrm{d} \sigma_\text{fid.}(AB \to AB \chi^+ \chi^-)}
       {\mathrm{d} \sqrt{s}}
  = \int\limits_{\max(\hat p_T, \tilde p_T)}
               ^{\frac{\sqrt{s}}{2} \sqrt{1 - \frac{4 m_\chi^2}{s}}}
    \mathrm{d} p_T
    \, \frac{\mathrm{d} \sigma(\gamma \gamma \to \chi^+ \chi^-)}{\mathrm{d} p_T}
    \, \frac{\mathrm{d} L_{AB}^\text{fid.}}{\mathrm{d} \sqrt{s}},
  \label{fiducial-xsection}
  \\
  \frac{\mathrm{d} L_{AB}^\text{fid.}}{\mathrm{d} \sqrt{s}}
  = \frac{\sqrt{s}}{2}
    \int\limits_{\max(-\hat y, \tilde y)}^{\min(\hat y, \tilde Y)}
    \mathrm{d} y
    \, n_A \left( \frac{\sqrt{s}}{2} \mathrm{e}^y \right)
    \, n_B \left( \frac{\sqrt{s}}{2} \mathrm{e}^{-y} \right),
\end{gather}
where $\tilde y$ and $\tilde Y$ are the constraints on rapidity coming from the
constraints on photon energies~\eqref{constraints-upc},
\begin{gather}
  \begin{aligned}
    \tilde y &= \max \left(
      \ln \frac{\hat \omega_{1,\text{min}}}{\sqrt{s} / 2},
      \ln \frac{\sqrt{s} / 2}{\hat \omega_{2,\text{max}}}
    \right),
    \\
    \tilde Y &= \min \left(
      \ln \frac{\hat \omega_{1,\text{max}}}{\sqrt{s} / 2},
      \ln \frac{\sqrt{s} / 2}{\hat \omega_{2,\text{min}}}
    \right),
  \end{aligned}
  \label{rapidity-constraints}
\end{gather}
and $\tilde p_T$ is an extra constraint on $p_T$ that ensures that integrations
are performed over physically meaningful domains. The role of the latter is to
select such values of $p_T$ that the following inequalities hold:
\begin{equation}
  \hat y > 0,\ -\hat y < \tilde Y,\ \hat y > \tilde y.
  \label{rapidity-sanity}
\end{equation}
Let
\begin{equation}
  P_T(Y)
  = \frac{\sqrt{s}}{4}
    \left[
        \frac{
            1
          + \sqrt{
                1
              - \frac{4 m_\chi^2}{s}
                \left( 1 + \frac{\sinh^2 Y}{\cosh^2 \hat \eta} \right)
            }
        }{\cosh(Y - \hat \eta)}
      - \frac{
            1
          - \sqrt{
                1
              - \frac{4 m_\chi^2}{s}
                \left( 1 + \frac{\sinh^2 Y}{\cosh^2 \hat \eta} \right)
            }
        }{\cosh(Y + \hat \eta)}
    \right].
\end{equation}
Then the solution of~\eqref{rapidity-sanity} is (assuming $\tilde y < \tilde Y$)
\begin{equation}
       p_T
  > \tilde p_T
  = \left\{
      \begin{aligned}
           &\max(P_T(0), P_T(\max(\tilde y, -\tilde Y)))
           &&\text{ if } \ 
               \frac{4 m_\chi^2}{s}
               \left(
                 1 + \frac{\sinh^2(\max(\tilde y, -\tilde Y))}{\cosh^2 \hat \eta}
               \right)
             < 1,
        \\ &P_T(0) &&\text{ otherwise,}
      \end{aligned}
    \right.
\end{equation}
where, for the reference,
\begin{equation}
  P_T(0) = \frac{\sqrt{s}}{2 \cosh \hat \eta} \sqrt{1 - \frac{4 m_\chi^2}{s}}.
\end{equation}

In the case when no constraints are imposed on the photons energies, i.e. when
$\hat \omega_{1,\text{min}} = \hat \omega_{2,\text{min}} = 0$, $\hat
\omega_{1,\text{max}} = \hat \omega_{2,\text{max}} = \infty$, the corresponding
limits on rapidity are $\tilde y = -\infty$, $\tilde Y = \infty$. Then the
integration limits are just $[-\hat y, \hat y]$, and $\tilde p_T = P_T(0)$. When
there are also no constraints on the particles transversal momentum and
pseudorapidity ($\hat p_T = 0$, $\hat \eta = \infty$), \eqref{fiducial-xsection}
reduces to~\eqref{xsection}.

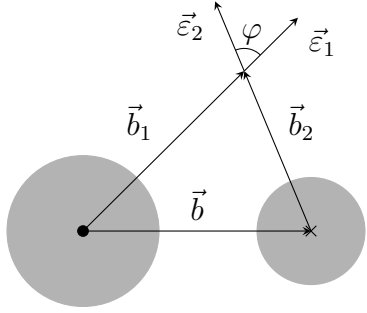
\begin{figure}[!tb]
  \centering
  \begin{tikzpicture}[>=stealth]
    \tikzmath{\R = 1; \l = 0.07;}

    \coordinate (Cl) at (-1.5 * \R, 0);
    \coordinate (Cr) at ( 1.5 * \R, 0);

    \filldraw [black!30!white] (Cl) circle [radius=\R];
    \filldraw (Cl) circle [radius=\l];
    \filldraw [black!30!white] (Cr) circle [radius=0.7\R];
    \draw     ($(Cr) + (\l, \l)$) -- ($(Cr) - (\l, \l)$);
    \draw     ($(Cr) + (\l, -\l)$) -- ($(Cr) + (-\l, \l)$);

    \draw [->] (Cl) -- node [above] {$\vec b$} (Cr);

    \coordinate (A) at ($(Cl) + (45:3 * \R)$);

    \draw [->] (Cl) -- node [above left]  {$\vec b_1$} (A);
    \draw [->] (Cr) -- node [above right] {$\vec b_2$} (A);

    \draw [->] (A) -- +(45:1)    node [below right] {$\vec \varepsilon_1$};
    \draw [->] (A) -- +(112.5:1) node [below left]  {$\vec \varepsilon_2$};
    \draw      ($(A) + (45:0.3)$) arc (45:112.5:0.3);
    \node at ($(A) + (78.75:0.5)$) {$\varphi$};
  \end{tikzpicture}
  \caption{
    An ultraperipheral collision of two particles moving perpendicular to the
    figure plane. $\vec b_1$ and $\vec b_2$ are impact parameters of a point in
    space at which photons collide relative to the corresponding particles. $b =
    \abs{\vec b_1 - \vec b_2}$ is the impact parameter of the collision.  $\vec
    \varepsilon_1$ and $\vec \varepsilon_2$ are the photons polarization
    vectors.
  }
  \label{f:upc}
\end{figure}

Let us now consider what happens if the colliding particles participate in weak
or strong interactions. If such interaction occurs, the final state of the
collision is not the same as in ultraperipheral collisions, so to calculate the
UPC cross section, we need to subtract the cross section for non-electromagnetic
interactions. Let $P_{AB}(b)$ be the probability for the colliding particles to
survive in a collision with the impact parameter $b$ (see Fig.~\ref{f:upc}).
The cross section can then be expressed as an integral over all possible
configurations of the three points in the collision plane~--- the centers of the
particles and the point of the photons collision (cf.~\eqref{epa-xsection}):
\begin{equation}
  \sigma(AB \to AB X)
  = \int\limits_0^\infty \mathrm{d} \omega_1
    \int\limits_0^\infty \mathrm{d} \omega_2
    \int \mathrm{d}^2 b_1
    \int \mathrm{d}^2 b_2
    \, \sigma(\gamma \gamma \to X)
    \, n_A(b_1, \omega_1)
    \, n_B(b_2, \omega_2)
    \, P_{AB}(b),
  \label{xsection-b-nn}
\end{equation}
where $\vec b_1$ and $\vec b_2$ are the position vectors of the photon-photon
collision point with respect to the centers of the particles $A$ and $B$, $b =
\abs{\vec b_1 - \vec b_2}$, $n(b_i, \omega)$ is the (spatial) spectrum of
equivalent photons at the distance $b_i$ from the source particle. The latter is
related to the spectrum $n(\omega)$~\eqref{epa-spectrum} through the equation
\begin{equation}
  n(\omega)
  = \int n(b, \omega) \, \mathrm{d}^2 b
  = 2 \pi \int\limits_0^\infty n(b, \omega) \, b \, \mathrm{d} b.
\end{equation}
Eq.~\eqref{xsection-b-nn} is derived under the assumption that the particles of
the system $X$ do not interact with $A$ or $B$ after the production.
Specifically, if $A$ or $B$ is a hadron, $X$ should not participate in strong
interactions; weak interactions and electromagnetic rescattering can usually be
neglected with the accuracy at the level of or better than 1\%.

Photons in ultraperipheral collisions are polarized as is shown in
Fig.~\ref{f:upc}, and $\sigma(\gamma \gamma \to X)$ in~\eqref{xsection-b-nn}
should take this polarization into account. Given cross sections for the
reactions with the polarizations of the photons parallel
($\sigma_\parallel(\gamma \gamma \to X)$) and perpendicular
($\sigma_\perp(\gamma \gamma \to X)$), this dependency can be factored
out~\cite{2106.14842}:
\begin{equation}
  \frac{\mathrm{d} \sigma(AB \to AB X)}{\mathrm{d} \sqrt{s}}
  = \sigma_\parallel(\gamma \gamma \to X)
    \frac{\mathrm{d} L_{AB}^\parallel}{\mathrm{d} \sqrt{s}}
  + \sigma_\perp(\gamma \gamma \to X)
    \frac{\mathrm{d} L_{AB}^\perp}{\mathrm{d} \sqrt{s}},
  \label{xsection-b}
\end{equation}
where
\begin{equation}
  \begin{aligned}
    \frac{\mathrm{d} L_{AB}^\parallel}{\mathrm{d} \sqrt{s}}
    &= \frac{\sqrt{s}}{2}
       \int \mathrm{d}^2 b_1
       \int \mathrm{d}^2 b_2
       \int\limits_{-\infty}^{\infty} \mathrm{d} y
       \, n_A \left( b_1, \frac{\sqrt{s}}{2} \mathrm{e}^y    \right)
       \, n_B \left( b_2, \frac{\sqrt{s}}{2} \mathrm{e}^{-y} \right)
       \, P_{AB}(b)
       \cos^2 \varphi,
    \\
    \frac{\mathrm{d} L_{AB}^\perp}{\mathrm{d} \sqrt{s}}
    &= \frac{\sqrt{s}}{2}
       \int \mathrm{d}^2 b_1
       \int \mathrm{d}^2 b_2
       \int\limits_{-\infty}^{\infty} \mathrm{d} y
       \, n_A \left( b_1, \frac{\sqrt{s}}{2} \mathrm{e}^y    \right)
       \, n_B \left( b_2, \frac{\sqrt{s}}{2} \mathrm{e}^{-y} \right)
       \, P_{AB}(b)
       \sin^2 \varphi
  \end{aligned}
  \label{luminosity-b}
\end{equation}
are the luminosities of photons with respective polarizations, $\varphi$ is the
angle between vectors $\vec b_1$, $\vec b_2$.

Calculation of the fiducial cross section with non-electromagnetic interactions
of the colliding particles taken into account follows the same approach as that
with non-electromagnetic interactions neglected~\eqref{fiducial-xsection}. The
result is
\begin{gather}
  \frac{\mathrm{d} \sigma_\text{fid.}(AB \to AB \chi^+ \chi^-)}
       {\mathrm{d} \sqrt{s}}
  = \hspace{-2ex}
    \int\limits_{\max(\hat p_T, \tilde p_T)}
               ^{\frac{\sqrt{s}}{2} \sqrt{1 - \frac{4 m_\chi^2}{s}}}
    \hspace{-2ex}
    \mathrm{d} p_T
    \left(
        \frac{\mathrm{d} \sigma_\parallel(\gamma \gamma \to \chi^+ \chi^-)}
             {\mathrm{d} p_T}
        \frac{\mathrm{d} L_{AB}^{\parallel,\text{fid.}}}{\mathrm{d} \sqrt{s}}
      + \frac{\mathrm{d} \sigma_\perp(\gamma \gamma \to \chi^+ \chi^-)}
             {\mathrm{d} p_T}
        \frac{\mathrm{d} L_{AB}^{\perp,\text{fid.}}}{\mathrm{d} \sqrt{s}}
    \right),
  \label{fiducial-xsection-b}
  \\
  \begin{aligned}
    \frac{\mathrm{d} L_{AB}^{\parallel,\text{fid.}}}{\mathrm{d} \sqrt{s}}
    &= \frac{\sqrt{s}}{2}
       \int \mathrm{d}^2 b_1
       \int \mathrm{d}^2 b_2
       \int\limits_{\max(-\hat y, \tilde y)}^{\min(\hat y, \tilde Y)}
       \mathrm{d} y
       \, n_A \left( b_1, \frac{\sqrt{s}}{2} \mathrm{e}^y    \right)
       \, n_B \left( b_2, \frac{\sqrt{s}}{2} \mathrm{e}^{-y} \right)
       \, P_{AB}(b)
       \, \cos^2 \varphi,
    \\
    \frac{\mathrm{d} L_{AB}^{\perp,\text{fid.}}}{\mathrm{d} \sqrt{s}}
    &= \frac{\sqrt{s}}{2}
       \int \mathrm{d}^2 b_1
       \int \mathrm{d}^2 b_2
       \int\limits_{\max(-\hat y, \tilde y)}^{\min(\hat y, \tilde Y)}
       \mathrm{d} y
       \, n_A \left( b_1, \frac{\sqrt{s}}{2} \mathrm{e}^y    \right)
       \, n_B \left( b_2, \frac{\sqrt{s}}{2} \mathrm{e}^{-y} \right)
       \, P_{AB}(b)
       \, \sin^2 \varphi.
    \label{luminosity-fid-b}
  \end{aligned}
\end{gather}

In the case of proton-proton collisions, the following expression can be used
for the probability to avoid non-electromagnetic
interactions~\cite{hep-ph/0608271}:
\begin{equation}
  P_{pp}(b) = \left( 1 - \mathrm{e}^{-\frac{b^2}{2 B}} \right)^2,
\end{equation}
where $B$ is an empirical parameter depending on the collision energy
$E$~\cite{1112.3243}:
\begin{equation}
  B = B_0 + 2 B_1 \ln(E / E_0) + 4 B_2 \ln^2(E / E_0),
  \label{pp-elastic-slope}
\end{equation}
$B_0 = 12~\text{GeV}^{-2}$, $B_1 = -0.22 \pm 0.17~\text{GeV}^{-2}$, $B_2 = 0.037
\pm 0.006~\text{GeV}^{-2}$, $E_0 = 1$~GeV. With this $P_{pp}(b)$, integration
over the angles in~\eqref{luminosity-b} (or~\eqref{luminosity-fid-b}~--- they
differ only in the limits of integration with respect to $y$) can be performed
analytically~\cite{2106.14842}:
\begin{equation}
  \begin{gathered}
    \begin{multlined}
      \frac{\mathrm{d} L_{pp}^{\parallel}}{\mathrm{d} \sqrt{s}}
      = \pi^2 \sqrt{s}
        \int\limits_0^\infty b_1 \, \mathrm{d} b_1
        \int\limits_0^\infty b_2 \, \mathrm{d} b_2
        \int\limits_{-\infty}^\infty \mathrm{d} y
        \, n_p \left( b_1, \tfrac{\sqrt{s}}{2} \, \mathrm{e}^y    \right)
        \, n_p \left( b_2, \tfrac{\sqrt{s}}{2} \, \mathrm{e}^{-y} \right)
        \\  \times
        \left\{
            1
          - 2 \mathrm{e}^{-\frac{b_1^2 + b_2^2}{2 B}}
            \left[
                I_0 \left( \frac{b_1 b_2}{B} \right)
              + I_2 \left( \frac{b_1 b_2}{B} \right)
            \right]
          + \mathrm{e}^{-\frac{b_1^2 + b_2^2}{B}}
            \left[
                I_0 \left( \frac{2 b_1 b_2}{B} \right)
              + I_2 \left( \frac{2 b_1 b_2}{B} \right)
            \right]
        \right\},
    \end{multlined}
    \\
    \begin{multlined}
      \frac{\mathrm{d} L_{pp}^{\perp}}{\mathrm{d} \sqrt{s}}
      = \pi^2 \sqrt{s}
        \int\limits_0^\infty b_1 \, \mathrm{d} b_1
        \int\limits_0^\infty b_2 \, \mathrm{d} b_2
        \int\limits_{-\infty}^\infty \mathrm{d} y
        \, n_p \left( b_1, \tfrac{\sqrt{s}}{2} \, \mathrm{e}^y    \right)
        \, n_p \left( b_2, \tfrac{\sqrt{s}}{2} \, \mathrm{e}^{-y} \right)
        \\  \times
        \left\{
            1
          - 2 \mathrm{e}^{-\frac{b_1^2 + b_2^2}{2 B}}
            \left[
                I_0 \left( \frac{b_1 b_2}{B} \right)
              - I_2 \left( \frac{b_1 b_2}{B} \right)
            \right]
          + \mathrm{e}^{-\frac{b_1^2 + b_2^2}{B}}
            \left[
                I_0 \left( \frac{2 b_1 b_2}{B} \right)
              - I_2 \left( \frac{2 b_1 b_2}{B} \right)
            \right]
        \right\},
    \end{multlined}
  \end{gathered}
  \label{pp-luminosity-b}
\end{equation}
where $n_p(b, \omega)$ is the equivalent photon spectrum of proton, $I_0(x)$,
$I_2(x)$ are the modified Bessel functions of the first kind.

\subsection{EPA spectra}

Equivalent photon spectrum of an electrically charged particle depends on its
internal structure. The structure is reflected in the electromagnetic current of
the particle through form factors. For an overview of proton electromagnetic
form factors, see, e.g.,~\cite{prep550-1}.

Electromagnetic current of a fermion:
\begin{equation}
  \mathcal{J}_\mu = Ze \bar \psi \left[
    F_1(Q^2) \gamma_\mu - \frac{\sigma_{\mu \nu} q^\nu}{2 m_\psi} F_2(Q^2)
  \right] \psi,
  \ 
  \sigma_{\mu \nu} = \frac{\gamma_\mu \gamma_\nu - \gamma_\nu \gamma_\mu}{2},
  \label{current}
\end{equation}
where $Ze$ is the fermion charge, $\psi$ is the fermion annihilation operator,
$\gamma_\mu$ are the Dirac matrices, $m_\psi$ is the fermion mass, $F_1(Q^2)$
and $F_2(Q^2)$ are the Dirac and Pauli fermion form factors, $Q^2$ is the photon
virtuality~\eqref{photon-virtuality}. The form factors are often expressed in
terms of the Sachs electric and magnetic form factors $G_E(Q^2)$,
$G_M(Q^2)$~\cite{pr126-2256}:
\begin{equation}
  \begin{aligned}
       G_M(Q^2) &= F_1(Q^2) + F_2(Q^2),
    &  F_1(Q^2)
       &= \frac{G_E(Q^2) + \frac{Q^2}{4 m_\psi^2} G_M(Q^2)}
               {1 + \frac{Q^2}{4 m_\psi^2}},
    \\ G_E(Q^2) &= F_1(Q^2) - \frac{Q^2}{4 m_\psi^2} F_2(Q^2),
    &  F_2(Q^2) &= \frac{G_M(Q^2) - G_E(Q^2)}{1 + \frac{Q^2}{4 m_\psi^2}}.
    \label{ff-sachs}
  \end{aligned}
\end{equation}

Form factors and the equivalent photon spectrum are related through the
following equation~\cite[Appendix~D]{prep15-181}
\begin{equation}
  n(\omega)
  = \frac{2 Z^2 \alpha}{\pi \omega}
    \int\limits_0^\infty
      \frac{D(q_\perp^2 + (\omega / \gamma)^2)}
           {(q_\perp^2 + (\omega / \gamma)^2)^2}
      \, q_\perp^3
      \, \mathrm{d} q_\perp,
  \label{spectrum}
\end{equation}
where
$\alpha = e^2 / 4 \pi$ is the fine structure constant, and
\begin{equation}
  D(Q^2) = \frac{G_E^2(Q^2) + \frac{Q^2}{4 m_\psi^2} G_M^2(Q^2)}
                {1 + \frac{Q^2}{4 m_\psi^2}}
  \label{ff-D}
\end{equation}
(cf. $F_1(Q^2)$ in~\eqref{ff-sachs}). The expression for the spectrum $n(b,
\omega)$ has not been derived yet.

If the Pauli form factor is neglected, i.e., $F_2(Q^2) = 0$, then
\begin{gather}
  n_E(\omega)
  = \frac{2 Z^2 \alpha}{\pi \omega}
    \int\limits_0^\infty
      \left[
        \frac{F_1(q_\perp^2 + (\omega / \gamma)^2)}
             {q_\perp^2 + (\omega / \gamma)^2}
      \right]^2
      q_\perp^3
    \, \mathrm{d} q_\perp,
  \label{spectrum-electric}
  \\
  n_E(b, \omega)
  = \frac{Z^2 \alpha}{\pi^2 \omega}
    \left[
      \int\limits_0^\infty
        \frac{F_1(q_\perp^2 + (\omega / \gamma)^2)}
             {q_\perp^2 + (\omega / \gamma)^2}
        J_1(b q_\perp)
        \, q_\perp^2
      \mathrm{d} q_\perp
    \right]^2,
  \label{spectrum-b-electric}
\end{gather}
where $J_1(x)$ is the Bessel function of the first kind.

If the charged particle has no internal structure, i.e., $F_1(Q^2) = 1$ and
$F_2(Q^2) = 0$, its equivalent photon spectrum $n_0(\omega)$ is logarithmically
divergent. If there is some cutoff on the photon momentum in the rest frame of
the particle, $\hat q$, then in the leading logarithmic approximation
\begin{equation}
  n_0(\omega)
  = \frac{2 Z^2 \alpha}{\pi \omega} \ln \frac{\hat q \gamma}{\omega}.
  \label{spectrum-0}
\end{equation}
This formula allows for analytical calculation of the cross
section~\cite{1806.07238} and can be used as a rough approximation or to
calculate the cross section for the production of less energetic particles.  The
value of $\hat q$ and the corresponding photon energy $\hat q \gamma$ are
related to the point in integration over $q_\perp$ when either the fermion
structurelessness approximation or the equivalent photon approximation breaks.
Ref.~\cite{landau-epa} considered production of electrons in a collision of
structureless particles and used $\hat q = m_e$ for the cutoff, where $m_e$ is
the electron mass. For proton, a good cutoff is $\hat q \sim \Lambda_\text{QCD}
\sim 0.2$~GeV since beyond that the photon carries enough momentum to break the
proton apart. If the particle form factor is known but the calculation of the
spectrum through~\eqref{spectrum}--\eqref{spectrum-b-electric} is undesirable
(too difficult or too slow), $\hat q$ can be estimated by comparing the low
energy limit of the spectrum with~\eqref{spectrum-0}. For example, for
\nuc{Pb}{208}, depending on the parameters of the form factor used, $\hat
q_\text{Pb}$ can be estimated as equal to 18 or 30~MeV~\cite[Appendix
A]{1806.07238}.

The spatial equivalent photon spectrum of a pointlike particle is
\begin{equation}
  n_0(b, \omega)
  = \frac{Z^2 \alpha \omega}{\pi^2 \gamma}
    \, K_1^2 \left( \frac{b \omega}{\gamma} \right),
\end{equation}
where $K_1(x)$ is the modified Bessel function of the second kind (the Macdonald
function).

Monopole approximation is often used for form factors of heavy nuclei. Under
this approximation the Dirac form factor is
\begin{equation}
  F_1(Q^2) = \frac{1}{1 + \frac{Q^2}{\Lambda^2}}
  \qquad \text{(monopole approximation)},
\end{equation}
and the Pauli form factor is neglected. The $\Lambda$ parameter is related to
the nucleus charge radius $R$ through the expression
\begin{equation}
  R^2 = -6 \lim\limits_{Q^2 \to 0} \frac{\mathrm{d} G_E(Q^2)}{\mathrm{d} Q^2}.
  \label{ff-radius}
\end{equation}
In this case the Sachs electric form factor $G_E(Q^2)$ is equal to the Dirac
form factor, so 
\begin{equation}
  \Lambda^2 = \frac{6}{R^2}.
\end{equation}
Substituting here
\begin{equation}
  R \approx (1.25~\text{fm}) \cdot A^{1/3},
\end{equation}
where $A$ is the number of nucleons in the nucleus, for \nuc{Pb}{208} we find
that $\Lambda \approx 70$~MeV.

Equivalent photon spectra under the monopole approximation:
\begin{gather}
  n_1(\omega)
  = \frac{Z^2 \alpha}{\pi \omega}
    \left[ (2 a + 1) \ln \left( 1 + \frac{1}{a} \right) - 2 \right],
  \ a = \left( \frac{\omega}{\Lambda \gamma} \right)^2,
  \\
  n_1(b, \omega)
  = \frac{Z^2 \alpha}{\pi^2 \omega}
    \left[
        \frac{\omega}{\gamma}
        K_1 \left( \frac{b \omega}{\gamma} \right)
      - \sqrt{\Lambda^2 + \left( \frac{\omega}{\gamma} \right)^2}
        K_1 \left(
          b \sqrt{\Lambda^2 + \left( \frac{\omega}{\gamma} \right)^2}
        \right)
    \right]^2.
\end{gather}

Dipole approximation is often used to describe form factors of proton. Under
this approximation the Sachs form factors are
\begin{equation}
  \begin{aligned}
    G_E(Q^2) &= \frac{1}{\left( 1 + \frac{Q^2}{\Lambda^2} \right)^2}, \\
    G_M(Q^2) &= \frac{\mu_\psi}{\left( 1 + \frac{Q^2}{\Lambda^2} \right)^2},
  \end{aligned}
  \qquad \left(
    \begin{array}{c}
      \text{dipole} \\ \text{approximation}
    \end{array}
  \right)
  \label{ff-dipole-sachs}
\end{equation}
where $\mu_\psi$ is the fermion magnetic moment expressed in units of $e / 2
m_\psi$, and $\Lambda$ is a parameter of the approximation related to the
fermion charge radius $R$ through~\eqref{ff-radius}:
\begin{equation}
  \Lambda^2 = \frac{12}{R^2}.
  \label{ff-dipole-lambda}
\end{equation}
For proton, the standard value of $\Lambda^2$ quoted in many textbooks is
$0.71~\text{GeV}^2$; however, it was obtained many years ago. Using the modern
value of $0.8414$~fm~\cite{rmp93-025010} for the proton charge radius,
via~\eqref{ff-dipole-lambda} we get that for proton $\Lambda^2 =
0.66~\text{GeV}^2$.

If the Pauli form factor is neglected, i.e., $F_2(Q^2) = 0$, then $F_1(Q^2) =
G_E(Q^2) = G_M(Q^2)$, and the equivalent photon spectra under the dipole
approximation are calculated through~\eqref{spectrum-electric},
\eqref{spectrum-b-electric}:
\begin{gather}
  n_2(\omega)
  = \frac{Z^2 \alpha}{\pi \omega}
    \left[
        (4 a + 1) \ln \left( 1 + \frac{1}{a} \right)
      - \frac{24 a^2 + 42 a + 17}{6 (a + 1)^2}
    \right],
  \ a = \left( \frac{\omega}{\Lambda \gamma} \right)^2,
  \label{spectrum-dipole}
  \\
  \begin{multlined}
    n_2(b, \omega)
    = \frac{Z^2 \alpha}{\pi^2 \omega}
      \left[
        \frac{\omega}{\gamma}
          K_1 \left( \frac{b \omega}{\gamma} \right)
        - \sqrt{\Lambda^2 + \left( \frac{\omega}{\gamma} \right)^2}
          K_1 \left( 
            b \sqrt{\Lambda^2 + \left( \frac{\omega}{\gamma} \right)^2}
          \right)
    \right. \\ \left.
        - \frac{b \Lambda^2}{2}
          K_0 \left(
            b \sqrt{\Lambda^2 + \left( \frac{\omega}{\gamma} \right)^2}
          \right)
       \right]^2.
  \end{multlined}
  \label{spectrum-dipole-b}
\end{gather}

If the term with the Pauli form factor is omitted from~\eqref{current}, i.e.,
$\mathcal{J}_\mu = Z e F_1(Q^2) \bar \psi \gamma_\mu \psi$, but the electric and
magnetic form factors are not assumed to be equal ($G_E(Q^2) \ne G_M(Q^2)$),
then the spectra calculated through~\eqref{spectrum-electric}
and~\eqref{spectrum-b-electric} are\footnote{
  The ``D'' subscript in $n_{2\text{D}}$ stands for ``Dirac''.
}
\begin{gather}
 \begin{split}
   n_{2\text{D}}(\omega)
   &= \frac{Z^2 \alpha}{\pi \omega}
      \left\{
          \left( 1 + 4 u - 2 (\mu_\psi - 1) \frac{u}{v} \right)
          \ln \left( 1 + \frac{1}{u} \right)
   \right. \\ &\qquad \left. {}
        + \frac{\mu_\psi - 1}{(v - 1)^4} \left[
              \frac{\mu_\psi - 1}{v - 1} (1 + 4 u + 3 v)
            - 2 \left( 1 + \frac{u}{v} \right)
          \right]
          \ln \frac{u + v}{u + 1}
        - \frac{24 u^2 + 42 u + 17}{6 (u + 1)^2}
   \right. \\ &\qquad \left. {}
        + (\mu_\psi - 1) \,
          \frac{
            6 u^2 (v^2 - 3 v + 3) + 3 u (3 v^2 - 9 v + 10) + 2 v^2 - 7 v + 11
          }{
            3 (u + 1)^2 (v - 1)^3
          }
   \right. \\ &\qquad \left. {}
        - (\mu_\psi - 1)^2 \,
          \frac{
            24 u^2 + 6 u (v + 7) - v^2 + 8 v + 17
          }{
            6 (u + 1)^2 (v - 1)^4
          }
      \right\},
  \end{split}
  \label{spectrum-dirac}
  \\
    \text{ where }
    u = \left( \frac{\omega}{\Lambda \gamma} \right)^2,
  \ v = \left( \frac{2 m_\psi}{\Lambda} \right)^2,
  \notag \\
  \begin{split}
    n_{2\text{D}}(b, \omega)
    &= \frac{Z^2 \alpha}{\pi^2 \omega}
       \left[
           \frac{\omega}{\gamma} K_1 \left( \frac{b \omega}{\gamma} \right)
         - \left(
               1
             + \frac{(\mu_\psi - 1) \frac{\Lambda^4}{16 m_\psi^4}}{
                 \left( 1 - \frac{\Lambda^2}{4 m_\psi^2} \right)^2
               }
           \right)
           \sqrt{\Lambda^2 + \frac{\omega^2}{\gamma^2}}
           \, K_1 \left( b \sqrt{\Lambda^2 + \frac{\omega^2}{\gamma^2}} \right)
    \right. \\ &\qquad \left.
         + \frac{(\mu_\psi - 1) \frac{\Lambda^4}{16 m_\psi^4}}{
             \left( 1 - \frac{\Lambda^2}{4 m_\psi^2}  \right)^2
           }
           \sqrt{4 m_\psi^2 + \frac{\omega^2}{\gamma^2}}
           \, K_1 \left( b \sqrt{4 m_\psi^2 + \frac{\omega^2}{\gamma^2}} \right)
    \right. \\ &\qquad \left.
         - \frac{1 - \frac{\mu_\psi \Lambda^2}{4 m_\psi^2}}
                {1 - \frac{\Lambda^2}{4 m_\psi^2}}
         \cdot
           \frac{b \Lambda^2}{2}
           \, K_0 \left( b \sqrt{\Lambda^2 + \frac{\omega^2}{\gamma^2}} \right)
       \right]^2.
  \end{split}
  \label{spectrum-dirac-b}
\end{gather}

Finally, with no assumptions on the second term in~\eqref{current} barring the
dipole approximation, the spectrum is calculated through~\eqref{spectrum}:
\begin{gather}
  \begin{multlined}
    n_p(\omega)
    = \frac{Z^2 \alpha}{\pi \omega}
      \left\{
        \left( 1 + 4 u - (\mu_\psi^2 - 1) \frac{u}{v} \right)
        \ln \left( 1 + \frac{1}{u} \right)
        - \frac{24 u^2 + 42 u + 17}{6 (u + 1)^2}
      \right. \\ \left.
        - \frac{\mu_\psi^2 - 1}{(v - 1)^3}
          \left[
              \frac{1 + u / v}{v - 1} \ln \frac{u + v}{u + 1}
            - \frac{6 u^2 (v^2 - 3v + 3) + 3u (3v^2 - 9v + 10) + 2v^2 - 7v + 11}
                   {6 (u + 1)^2}
          \right]
      \right\},
  \end{multlined}
  \label{spectrum-proton}
  \\
    u = \left( \frac{\omega}{\Lambda \gamma} \right)^2,
  \ v = \left( \frac{2 m_\psi}{\Lambda} \right)^2.
  \notag
\end{gather}
This is the correct spectrum for proton, however its spatial counterpart has not
been derived yet.

\subsection{Photons fusion}

Only cross sections for the production of a pair of fermions~\cite{pr46-1087}
are currently available in \texttt{libepa}:
\begin{align}
  \sigma_\parallel(\gamma \gamma \to \chi^+ \chi^-)
  &= \frac{4 \pi \alpha^2}{s}
     \left[
         \left(
           1 + \frac{4 m_\chi^2}{s} - \frac{12 m_\chi^4}{s^2}
        \right)
         \ln \frac{1 + \sqrt{1 - 4 m_\chi^2 / s}}{1 - \sqrt{1 - 4 m_\chi^2 / s}}
       - \left( 1 + \frac{6 m_\chi^2}{s} \right)
         \sqrt{1 - \frac{4 m_\chi^2}{s}}
     \right],
  \label{photons-to-fermions-parallel}
  \\
  \sigma_\perp(\gamma \gamma \to \chi^+ \chi^-)
  &= \frac{4 \pi \alpha^2}{s}
     \left[
         \left(
           1 + \frac{4 m_\chi^2}{s} - \frac{4 m_\chi^4}{s^2}
        \right)
         \ln \frac{1 + \sqrt{1 - 4 m_\chi^2 / s}}{1 - \sqrt{1 - 4 m_\chi^2 / s}}
       - \left( 1 + \frac{2 m_\chi^2}{s} \right)
         \sqrt{1 - \frac{4 m_\chi^2}{s}}
     \right],
  \label{photons-to-fermions-perp}
  \\
  \sigma(\gamma \gamma \to \chi^+ \chi^-)
  &= \frac{4 \pi \alpha^2}{s}
     \left[
         \left(
           1 + \frac{4 m_\chi^2}{s} - \frac{8 m_\chi^4}{s^2}
        \right)
         \ln \frac{1 + \sqrt{1 - 4 m_\chi^2 / s}}{1 - \sqrt{1 - 4 m_\chi^2 / s}}
       - \left( 1 + \frac{4 m_\chi^2}{s} \right)
         \sqrt{1 - \frac{4 m_\chi^2}{s}}
     \right],
  \label{photons-to-fermions}
\end{align}
\begin{align}
  \frac{\mathrm{d} \sigma_\parallel(\gamma \gamma \to \chi^+ \chi^-)}
       {\mathrm{d} p_T}
  &= \frac{8 \pi \alpha^2 p_T}{s (p_T^2 + m_\chi^2)}
     \cdot \frac{1 - \frac{2 (p_T^4 + 2 m_\chi^4)}{s (p_T^2 + m_\chi^2)}}
                {\sqrt{1 - \frac{4 (p_T^2 + m_\chi^2)}{s}}},
  \label{photons-to-fermions-parallel-pT} \\
  \frac{\mathrm{d} \sigma_\perp(\gamma \gamma \to \chi^+ \chi^-)}
       {\mathrm{d} p_T}
  &= \frac{8 \pi \alpha^2 p_T}{s (p_T^2 + m_\chi^2)}
     \cdot \frac{1 - \frac{2 p_T^4}{s (p_T^2 + m_\chi^2)}}
                {\sqrt{1 - \frac{4 (p_T^2 + m_\chi^2)}{s}}},
  \label{photons-to-fermions-perp-pT} \\
  \frac{\mathrm{d} \sigma(\gamma \gamma \to \chi^+ \chi^-)}
       {\mathrm{d} p_T}
  &= \frac{8 \pi \alpha^2 p_T}{s (p_T^2 + m_\chi^2)}
     \cdot \frac{1 - \frac{2 (p_T^4 + m_\chi^4)}{s (p_T^2 + m_\chi^2)}}
                {\sqrt{1 - \frac{4 (p_T^2 + m_\chi^2)}{s}}},
  \label{photons-to-fermions-pT}
\end{align}
where $m_\chi$ is the fermion mass.

\section{Library overview}

\label{s:library}

\texttt{libepa} is designed in the function programming paradigm. The central
concept of functional programming is function as a value. That includes storing
a function in a variable and having functions that accept and\slash{}or return
functions --- the latter are called \emph{higher-order functions}. Languages
supporting functional programming paradigm usually provide an operator that
creates anonymous functions, often also called \emph{lambda
functions}.\footnote{The ``lambda'' here is a reference to lambda calculus, the
mathematical model of computation introduced in 1930s~\cite{church}.} Depending
on the language, a function created in a context with some variables visible can
refer to these variables and modify them during its execution. Such a bundle of
a function and external variables it refers to (its environment) is called a
\emph{closure}.

\texttt{libepa} provides a set of C++ and Python higher-order functions that
allows the user of the library to construct functions (closures) representing
mathematical functions described in Section~\ref{s:physics}.  For example, to
calculate the differential cross section~\eqref{xsection} for the production of
a pair of muons with the invariant mass 100~GeV in collisions of protons with
the energy 13~TeV, the following code can be used
\begin{lstlisting}[style=c++]
const double muon_mass = 105.6583745e-3; // GeV
const double collision_energy = 13e3; // GeV
const double invariant_mass = 100; // GeV
auto luminosity = pp_luminosity(collision_energy);
auto photons_to_muons = photons_to_fermions(muon_mass);
auto cross_section = xsection(photons_to_muons, luminosity);
double result = cross_section(invariant_mass); // barn/GeV
\end{lstlisting}
Here \verb|pp_luminosity| is a function that takes the collision energy $E$ and
returns a closure --- a function that takes the invariant mass of the photon
pair $\sqrt{s}$ as its sole argument and returns the value of the photon-photon
luminosity in proton-proton collisions with the energy $E$ for the given
$\sqrt{s}$, $\frac{\mathrm{d} L_{pp}}{\mathrm{d} \sqrt{s}}(\sqrt{s})$. We can
also just say that the closure \emph{is} the photon-photon luminosity in
proton-proton collisions with the energy $E$, because this definition leaves
only the $s$ parameter undefined in~\eqref{luminosity} which then has to be
computable from the closure arguments. In this sense,
\verb|photons_to_fermions| returns the cross section for the production of a
pair of fermions with the mass equal to
\verb|muon_mass|~\eqref{photons-to-fermions}, and \verb|xsection| returns the
desired cross section~\eqref{xsection}.

The function \verb|xsection|~\eqref{xsection} would be quite straightforward to
implement in Python, and it may be instructive to do so here:
\begin{lstlisting}[style=python]
def xsection(photon_photon, luminosity):
  def closure(rs):
    return photon_photon(rs) * luminosity(rs)
  return closure
\end{lstlisting}
Here \verb|rs| is $\sqrt{s}$, the invariant mass of the system $X$.  Note how
\verb|xsection| takes a pair of functions and returns a function.

Many formulas listed in Section~\ref{s:physics} involve integrations, so to
compute them an efficient algorithm for numerical one-dimensional integration is
required. The default algorithm used by \texttt{libepa} is the adaptive
Gauss-Kronrod integration algorithm (QAG) provided by the GNU Scientific
Library (GSL)~\cite{gsl}. However, \texttt{libepa} was designed so that the user
is free to use their own integration algorithms. To facilitate that, each
function computing an integral takes an \emph{integrator} or an \emph{integrator
generator} as an optional argument. An integrator is a function computing
$\int_a^b f(x) \mathrm{d} x$, where $f(x)$, $a$ and $b$ are the integrator
arguments; $a$ or $b$ or both can be infinite. An integrator generator is a
function that takes an \emph{integration level} (an unnegative integer) and
returns an integrator. Integration level of an integral is the nesting number of
the integral in the calculation. Calculation of multiple integrals requires
higher numerical accuracy for the inner integrals, so the integrator generator
should use the provided integration level to tune the accuracy of the
integration algorithm. Whenever a \texttt{libepa} function takes an integrator
generator, it also accepts the initial integration level as an argument.

The accuracy of the GSL integration algorithms is governed by two variables,
\verb|absolute_error| and \verb|relative_error|. Integration stops when either
the absolute or the relative difference in the estimation of the integral
between successive integrations is less than the corresponding parameter.
Correct utilization of \verb|absolute_error| requires estimation of the
magnitude of the integral, so it is not used by default. The default value of
\verb|relative_error| at integration level $n$ is $10^{-3 - n}$, meaning that
the calculation accuracy should not be much worse than $0.1$\% at integration
level 0 and it improves by an order of magnitude for each subsequent integration
level.

Note one problem that frequently arises when multiple integrals are evaluated via
successive application of one-dimensional integration. Consider the integral
$\iint \mathrm{d} x \, \mathrm{d} y \, f(x, y)$ and suppose that when evaluating
the integral with respect to $y$ for two successive points $x_0$ and $x_1$, the
integration algorithm decides to increase the number of iterations.  It may
result in a bump in the function $\int \mathrm{d} y f(x, y)$ as perceived by the
integration algorithm integrating with respect to $x$. Unfortunately, it
violates the requirement that the integrand function has to be smooth that most
integration algorithms have, including QAG. In this case the algorithm may
proceed by increasing the number of points around the bump until satisfied or
until giving up due to roundoff errors. This situation may be mitigated by
tweaking the integration accuracy of both integrators, or by using an algorithm
that can handle irregularities of the integrand~--- CQUAD from the GSL may be a
good try.

\texttt{libepa} is written in C++ and provides bindings for Python. The bindings
are implemented through the C foreign function interface (CFFI), and having that
interface it should be easy to create bindings for other languages if needed.
The Python API is designed so that it can transparently work with both the
functions defined in Python and the functions written in C++ and returned by
\texttt{libepa}. Specifically, the functions returned by \texttt{libepa} can be
treated as ordinary Python functions, and the functions written in Python can be
used as arguments to \texttt{libepa} higher-order functions. While composing the
functions, \texttt{libepa} can detect when they were implemented in C++ and
optimize away the intermediate calls through Python. Thus, when the computation
function was built only from the functions returned by \texttt{libepa} (no user
functions written in Python were used), the speed of computations is the same in
Python as it is in a pure C++ program. 

To understand how this optimization is achieved, let us look at implementation
of closures in C++ and C. In C++, a closure is represented with the standard
function, \verb|std::function|, usually wrapping an anonymous class created by
the C++ lambda function operator \verb|[](){}|. In \texttt{libepa}
CFFI, a closure is the following structure:
\begin{lstlisting}[style=c++]
struct epa_function {
  void (*function)();
  void* data;
  void (*destructor)(void*);
}
\end{lstlisting}
Here \verb|function| is a pointer to a C function which is actually a machine
code stored in memory, and \verb|data| is a pointer to the closure environment
that contains the variables captured by the closure. \verb|data| is passed as
the last argument to \verb|function| and it is expected that \verb|function|
knows how to extract the variables from it. Finally, \verb|destructor| is a
pointer to a C function that frees the memory \verb|data| points to.  When a C++
\verb|std::function| is passed from \texttt{libepa} to CFFI (\textit{lifted}), a
\verb|epa_function| is created with \verb|data| pointing to the
\verb|std::function| and \verb|function| pointing to a \textit{trampoline}, a
C function that calls its last argument passing it the rest of the arguments and
catches the errors. When a \verb|epa_function| is received by \texttt{libepa}
from CFFI (\textit{lowered}), its \verb|function| pointer is compared to that of
the trampoline, and if equal, \verb|data| is casted to the pointer to the
appropriate \verb|std::function| and used as is, avoiding the intermediate call
to \verb|function|. 

Detailed description of \texttt{libepa} functions and data types can be found in
\texttt{libepa} documentation~\cite{libepa}.

\section{Examples}

\label{s:examples}

In this Section we provide some examples of how to use \texttt{libepa}. These
are the simplified versions of the examples that can be found in
\texttt{examples} directory in \texttt{libepa} sources. We concentrate on the
code working with \texttt{libepa} and give some hopefully useful comments on the
calculations.

\subsection{Generic integral calculation}
\label{s:integration}

Though \texttt{libepa} is designed for particular physics applications, parts of
its functionality can be used for calculation of generic multiple integrals. We
demonstrate here the approach used in the library internal functions by
calculating 
\begin{equation}
  I(a)
  \equiv \frac{15}{a}
         \int\limits_0^a \mathrm{d} x
      \, x
         \int\limits_0^{\sqrt{1 - \left( \frac{x}{a} \right)^2}}
         \mathrm{d} y
      \, y
         \int\limits_0^{\sqrt{1 - \left( \frac{x}{a} \right)^2 - y^2}}
         \mathrm{d} z
      \, \frac{z}{\sqrt{x^2 + y^2 + z^2}}
  =      \frac{a (a + \frac{1}{2})}{(a + 1)^2}.
  \label{integral}
\end{equation}
The following code performs the calculation for a set of points and prints the
result on the standard output:
\begin{lstlisting}[style=c++]
#include <epa/epa.hpp>

int main(void) {
  // build the computation function
  auto I = [
    // precompute integrators
    integrate_x = epa::default_integrator(0),
    integrate_y = epa::default_integrator(1),
    integrate_z = epa::default_integrator(2)
  ](double a) -> double {
    return 15.0 / a * integrate_x([&](double x) -> double {
        return x * integrate_y([&](double y) -> double {
            return y * integrate_z([&](double z) -> double {
                return z / sqrt(x*x + y*y + z*z);},
                0, sqrt(1 - pow(x / a, 2) - y * y));},
            0, sqrt(1 - pow(x / a, 2)));},
        0, a);
  };

  for (double a = 1; a <= 100; a += 1)
    printf("%3.0f\t%.5f\t%.5f\n", a, I(a), a * (a + 0.5) / pow(a + 1, 2));
  return 0;
};
\end{lstlisting}
Here \verb|I| is a closure computing $I(a)$. It captures variables
\verb|integrate_x|, \verb|integrate_y| and \verb|integrate_z| which are
integrators used to calculate the integrals with respect to $x$, $y$ and $z$
respectively. For efficiency (although not important in this simple example), it
is essential that they are created once along with the closure rather than at
every call to \verb|I| or, worse, at every call to each of the inner lambda
functions.

Python interface is even easier to use due to garbage collection available in
Python. However, in this case all inner functions are calculated by Python, and
it appears that the cost of crossing the C++/Python interface is higher in
\texttt{libepa} than in \texttt{SciPy}~\cite{scipy}, the Python library
implementing the same integration algorithms as GSL.\footnote{
  GSL and \texttt{SciPy} use different implementations of the same integration
  functions originally implemented in QUADPACK~\cite{quadpack}. \texttt{SciPy}
  \texttt{quad} integration algorithm with the default parameters is the same as
  that of GSL \texttt{gsl\_integration\_qag} function with \texttt{key} set to
  \texttt{GAUSS21}.
}

\subsection{Cross section calculation}

\label{s:xsection}

Let us plot the cross section for the production of a pair of fermions in
ultraperipheral proton-proton collisions with the energy $E = 13$~TeV for the
fermion mass range from 90 to 250 GeV. To do so, we need to calculate
\begin{equation}
  \sigma(pp \to pp \chi^+ \chi^-)
  = \int\limits_{2 m_\chi}^{E / 2}
      \frac{\mathrm{d} \sigma(pp \to pp \chi^+ \chi^-)}{\mathrm{d} \sqrt{s}}
    \mathrm{d} \sqrt{s},
  \label{example-xsection}
\end{equation}
for a number of fermion masses $m_\chi$ spanning the desired range. The
integrand is defined by~\eqref{xsection} with $A = B = p$, $X = \chi^+ \chi^-$. 

First, let us compute the cross section neglecting non-electromagnetic
interactions. Here is the Python code that does the calculation and prints on
the standard output two columns of numbers: $m_\chi$ and the corresponding cross
section:
\begin{lstlisting}[style=python]
import epa

integrate = epa.default_integrator(0)
luminosity = epa.pp_luminosity(13e3, integrator = epa.default_integrator(1))
def xsection(mass):
    return integrate(
            epa.xsection(epa.photons_to_fermions(mass), luminosity),
            2 * mass,
            6.5e3
    )
for mass in range(90, 251, 5):
    print(f'{mass:3d} {xsection(mass):19.12e}')
\end{lstlisting}
The code begins with calling \verb|default_integrator| with integration level 0.
The returned closure stored in \verb|integrate| is the integrator that will be
used for the outermost integration~\eqref{example-xsection}.
\verb|pp_luminosity| returns the photon-photon luminosity in proton-proton
collisions~\eqref{luminosity} with the photon spectrum defined
by~\eqref{spectrum-proton}. The \verb|13e3| is the collision energy in GeV; in
our problem the integral in~\eqref{luminosity} has integration level 1, so we
need to provide the integrator explicitly. Next we define the cross section as a
function of $m_\chi$. \verb|photons_to_fermions| returns the cross section for
the production of a fermion pair in a fusion of
photons~\eqref{photons-to-fermions}, \verb|xsection| returns the differential
cross section~\eqref{xsection}. The latter is integrated by \verb|integrate|.
The final two lines of the code iterate across the mass range, call
\verb|xsection| and print its result.

Now, let us take into account non-electromagnetic interactions:
\begin{lstlisting}[style=python]
import epa
import os
import sys
import threading

queue_mutex = threading.Lock()
result_mutex = threading.Lock()
masses = iter(range(90, 251, 5))
result = []

epa.set_default_relative_error(1e-2)

def work():
    integrate = epa.default_integrator(0)
    luminosity = epa.pp_luminosity_b(13e3, integration_level = 1)
    def xsection(mass):
        return integrate(
                epa.xsection_b(epa.photons_to_fermions_b(mass), luminosity),
                2 * mass,
                3e3
        )
    while True:
        with queue_mutex:
            try:
                mass = next(masses)
            except StopIteration:
                return
        value = xsection(mass)
        with result_mutex:
            result.append((mass, value))
            print(f'{mass:3d} {value:19.12e}', file = sys.stderr)

threads = []
for cpu in os.sched_getaffinity(0):
    thread = threading.Thread(target = work, daemon = True)
    thread.start()
    threads.append(thread)
for thread in threads:
    thread.join()

for mass, xsection in sorted(result, key = lambda x: x[0]):
    print(f'{mass:3d} {xsection:19.12e}')
\end{lstlisting}
In this case we need to calculate a quadruple integral defined
by~\eqref{xsection-b} and~\eqref{pp-luminosity-b}. This calculation takes time,
so it is parallelized: cross sections for several fermion masses are computed at
once.  The tricky part here is that the integration functions are not thread
safe since they use the workspace allocated at the integrator creation time,
during the calls to higher-order functions composing the computation function.
That means that the computation function has to be constructed individually for
each thread, so it is done in the thread \verb|work| function in lines 14--21.
The difference to the previous listing are the \verb|_b| suffixes to function
names. These refer to the functions dealing with non-electromagnetic
interactions (integrating over $\vec b_1$, $\vec b_2$ in Fig.~\ref{f:upc}). The
function \verb|photons_to_fermions_b| returns a closure that returns two values
at each call: the cross sections~\eqref{photons-to-fermions-parallel} and
\eqref{photons-to-fermions-perp}. The function returned by \verb|xsection_b|
passes these cross sections as a pair to \verb|luminosity| to compute the cross
section for the ultraperipheral collision~\eqref{xsection-b}. Refer to the
documentation of \verb|luminosity_b| for details.

Recall from Section~\ref{s:library} that by default the accuracy of the
calculation of the integral with integration level 4 is $10^{-7}$. This is quite
a strong requirement resulting in roundoff errors raised by the GSL QAG
integration algorithm. To avoid those, the accuracy of the calculation was
reduced from $\sim 0.1$\% to $\sim 1$\% in line~11. Also, since it is difficult
to calculate the differential cross section at high invariant masses where there
are few photons that have enough energy to produce the fermion pair, the upper
integration limit was decreased to 3~TeV in line~20.

The rest of the code setups a multithreading computation and collects the
results; we leave it for the reader to understand how it works. Note that all
computations are performed by C++ functions, so Python's global interpreter lock
does not get in the way here.

\begin{figure}[!tb]
  \includegraphics{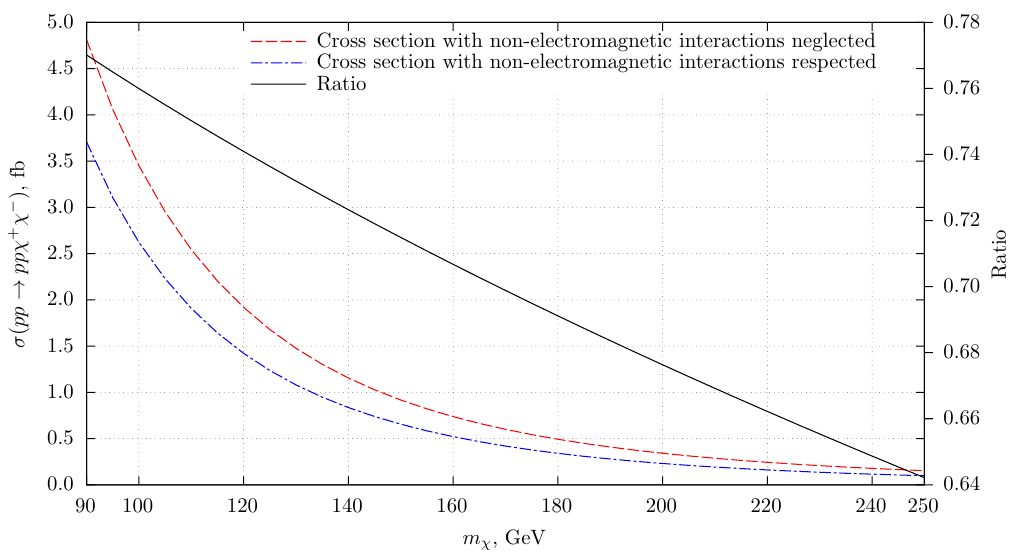}
  \caption{
    Cross section for the production of a fermion pair in ultraperipheral
    collisions of protons with the energy 13~TeV depending on the fermion mass.
  }
  \label{f:xsection}
\end{figure}

Results of the calculation are presented in Fig.~\ref{f:xsection}.

\subsection{Survival factor}

\label{s:survival}

Let us now calculate photon-photon luminosities in proton-proton collisions with
and without taking into account non-electromagnetic interactions. Their ratio is
called the survival factor. It can be calculated once and then used in various
problems to estimate the necessity to take into account non-electromagnetic
interactions. We consider three equivalent photon spectra for the proton: no
neglections~\eqref{spectrum-proton}, with the the second term in
current~\eqref{current} omitted but distinct electric and magnetic form
factors~\eqref{spectrum-dirac}, \eqref{spectrum-dirac-b}, and with the Pauli
form factor neglected properly~\eqref{spectrum-dipole},
\eqref{spectrum-dipole-b}.  Notations, approximations and equations used in
the calculation are summarized in Table~\ref{t:survival-notation}. Note that the
spatial equivalent photon spectra for the proton $n_p(b, \omega)$ that would
take into account the Pauli form factor has not been derived yet, so the
corresponding luminosity (and the survival factor) cannot be calculated.

\begin{table}[!tb]
  \centering
  \caption{Notation and description of the luminosities computed in
  section~\ref{s:survival}.}
  \begin{tabular}{r|ccccc}
    Notation
    & $\tilde L$
    & $L_{2\text{D}}$
    & $\tilde L_{2 \text{D}}$
    & $L_2$
    & $\tilde L_2$
    \\ \hline
    Non-electromagnetic interactions
    & no
    & yes
    & no
    & yes
    & no
    \\
    Second term in~\eqref{current}
    & yes
    & no
    & no
    & no
    & no
    \\
    Electric and magnetic form factors
    & distinct
    & distinct
    & distinct
    & equal
    & equal
    \\
    Formula
    & \eqref{luminosity}
    & \eqref{pp-luminosity-b}$^\dagger$
    & \eqref{luminosity}
    & \eqref{pp-luminosity-b}$^\dagger$
    & \eqref{luminosity}
    \\
    Spectrum
    & \eqref{spectrum-proton}
    & \eqref{spectrum-dirac-b}
    & \eqref{spectrum-dirac}
    & \eqref{spectrum-dipole-b}
    & \eqref{spectrum-dipole}
    \\
    Survival factor
    &
    & \multicolumn{2}{c}{
      $
        S_{2\text{D}}
        = \frac{\mathrm{d} L_{2\text{D}} / \mathrm{d} \sqrt{s}}
               {\mathrm{d} \tilde L_{2\text{D}} / \mathrm{d} \sqrt{s}}
      $
    }
    & \multicolumn{2}{c}{
      $
        S_2
        = \frac{\mathrm{d} L_2 / \mathrm{d} \sqrt{s}}
               {\mathrm{d} \tilde L_2 / \mathrm{d} \sqrt{s}}
      $
    }
    \\
    \multicolumn{6}{l}{\footnotesize $^\dagger$ summed over polarizations.}
  \end{tabular}
  \label{t:survival-notation}
\end{table}

The spectrum~\eqref{spectrum-proton} is what the function \verb|pp_luminosity|
uses, so $\mathrm{d} \tilde L / \mathrm{d} \sqrt{s}$ is calculated with just a
couple of lines of code:
\begin{lstlisting}[style=python]
import epa

luminosity = epa.pp_luminosity(13e3)
for rs in range(1, 1000):
    print(f'{rs:4d} {luminosity(rs):19.12e}')
\end{lstlisting}

The spectrum~\eqref{spectrum-dirac-b} is what the function
\verb|pp_luminosity_b| uses, so calculation of $\mathrm{d} L_{2\text{D}}
/ \mathrm{d} \sqrt{s}$ is also a simple matter:\footnote{
  This calculation takes a lot of time, as is typical for calculations with
  non-electromagnetic interactions taken into account. It can and should be
  parallelized using the technique shown in Section~\ref{s:xsection} or similar.
  In the following we concentrate on the code working with \texttt{libepa} and
  do not mention possible general optimizations. Complete code can be found in
  the supplementary materials.
  \label{fn:parallelization}
}
\begin{lstlisting}[style=python]
import epa

luminosity = epa.pp_luminosity_b(13e3)
for rs in range(1, 1000):
    print(f'{rs:4d} {luminosity(rs, (1, 1)):19.12e}')
\end{lstlisting}
Note the extra argument \verb|(1, 1)| to \verb|luminosity| in the last line.
Luminosity with non-electromagnetic interactions taken into account depends on
the photons polarizations. Computation of the luminosity takes a lot of time,
but luminosities for parallel and perpendicular photon polarizations can be
applied simultaneously. To facilitate that, functions computing luminosities
expect an extra argument which should be a tuple with the values being the
photons parallel $p_\parallel$ and perpendicular $p_\perp$ polarizations
respectively. The value computed is then
\begin{equation}
  \frac{\mathrm{d} L_{AB}}{\mathrm {d} \sqrt{s}}(p_\parallel, p_\perp)
  = p_\parallel \frac{\mathrm{d} L_{AB}^\parallel}{\mathrm{d} \sqrt{s}}
  + p_\perp \frac{\mathrm{d} L_{AB}^\perp}{\mathrm{d} \sqrt{s}},
\end{equation}
where $\mathrm{d} L_{AB}^\parallel / \mathrm{d} \sqrt{s}$, $\mathrm{d}
L_{AB}^\perp / \mathrm{d} \sqrt{s}$ are the luminosities~\eqref{luminosity-b}.
To calculate the luminosity for polarized photons, pass \verb|(1, 0)| or
\verb|(0, 1)|; to sum or average over the polarizations, pass \verb|(1, 1)| or
\verb|(0.5, 0.5)| respectively; to calculate the UPC cross section, pass the
photon fusion cross sections according to~\eqref{xsection-b}. For the details,
see the documentation on the type \verb|Luminosity_b| and the function
\verb|luminosity_b|.

To calculate $\mathrm{d} \tilde L_{2\text{D}} / \mathrm{d} \sqrt{s}$, just call
\verb|luminosity| with the desired spectrum:
\begin{lstlisting}[style=python]
import epa

spectrum = epa.proton_dipole_spectrum_Dirac(6.5e3)
luminosity = epa.luminosity(spectrum)
for rs in range(1, 1000):
    print(f'{rs:4d} {luminosity(rs):19.12e}')
\end{lstlisting}
Here the \verb|6.5e3| is the proton energy in GeV.

For $\mathrm{d} L_2 / \mathrm{d} \sqrt{s}$:
\begin{lstlisting}[style=python]
import epa

lorentz    = 6.5e3 / epa.proton_mass
lambda2    = epa.proton_dipole_form_factor_lambda2
spectrum   = epa.spectrum_dipole(1, lorentz, lambda2)
spectrum_b = epa.spectrum_b_dipole(1, lorentz, lambda2)
luminosity = epa.ppx_luminosity_b(
        spectrum, spectrum_b, epa.pp_elastic_slope(13e3)
)
for rs in range(1, 1000):
    print(f'{rs:4d} {luminosity(rs, (1, 1)):19.12e}')
\end{lstlisting}
Here \verb|lorentz| is the proton Lorentz factor $\gamma$, \verb|lambda2| is the
form factor parameter $\Lambda^2$~\eqref{ff-dipole-lambda}, \verb|spectrum| is
the spectrum~\eqref{spectrum-dipole}, \verb|spectrum_b| is its spatial
counterpart~\eqref{spectrum-dipole-b}. \verb|ppx_luminosity_b| returns a
function calculating~\eqref{pp-luminosity-b} with the proton spectrum $n_p(b,
\omega)$ replaced with its second argument, \verb|spectrum_b| in this code. The
first argument, \verb|spectrum|, is optional, and when provided it is used to
calculate the first term in the braces (the 1) in~\eqref{pp-luminosity-b} --- it
speeds up calculations and helps to avoid roundoff errors when subtracting small
values from the 1.  \verb|pp_elastic_slope| calculates the value of
$B$~\eqref{pp-elastic-slope}.

Finally, for $\mathrm{d} \tilde L_2 / \mathrm{d} \sqrt{s}$, we could also use
the \verb|spectrum_dipole| function, but let us pretend that it does not exist
and show how one could implement such a function in Python:
\begin{lstlisting}[style=python]
import epa
import math

def spectrum_dipole(Z, lorentz, lambda2):
    c = Z**2 * epa.alpha / math.pi
    def spectrum(energy):
        a = (energy / lorentz)**2 / lambda2
        return c / energy * (
                  (1 + 4 * a) * math.log(1 + 1 / a)
                - ((24 * a + 42) * a + 17) / (6 * (a + 1)**2)
        )
    return spectrum

lorentz = 6.5e3 / epa.proton_mass
lambda2 = epa.proton_dipole_form_factor_lambda2
spectrum = spectrum_dipole(1, lorentz, lambda2)
luminosity = epa.luminosity(spectrum)

for rs in range(1, 1000):
    print(f'{rs:4d} {luminosity(rs):19.12e}')
\end{lstlisting}

\begin{figure}[!tb]
  \centering
  \includegraphics{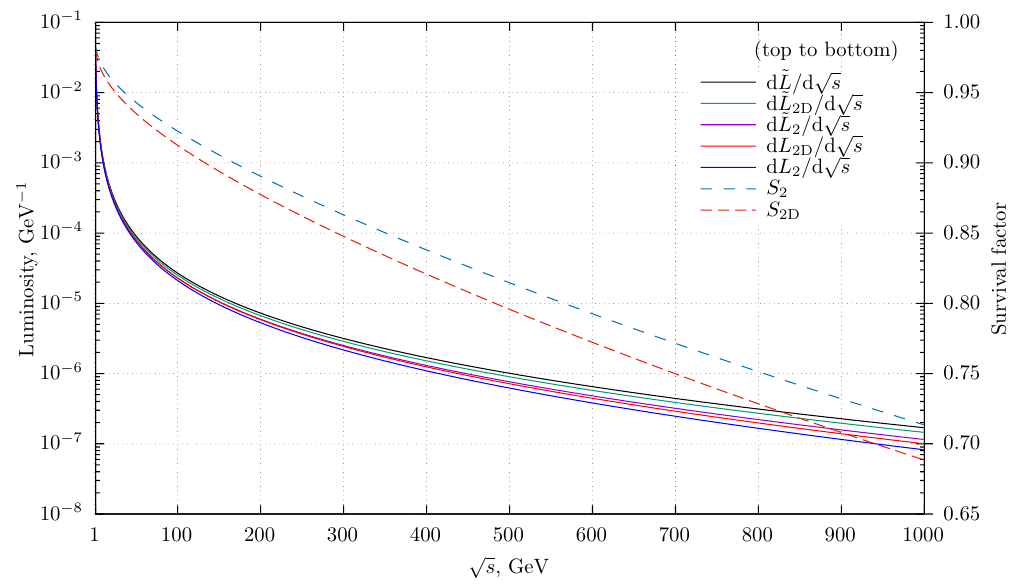}
  \caption{
    Photon-photon luminosity in ultraperipheral collisions of protons with the
    energy 13~TeV under various approximations described in
    Table~\ref{t:survival-notation}.
  }
  \label{f:survival}
\end{figure}

Results of the calculations can be found in Fig.~\ref{f:survival}.

\subsection{Muon pair production at the LHC}

\label{s:example-atlas}

The ATLAS Collaboration has measured muon pair production cross section in
ultraperipheral proton-proton collisions at the LHC~\cite{1708.04053}. We have
already considered this problem in~\cite{1806.07238, 2106.14842}, so let us here
just briefly describe the calculation and show the code and the results with the
updated values of physical constants. Since the spatial equivalent photon
spectrum for proton $n_p(b, \omega)$ is unavailable, we calculate three cross
sections $\tilde \sigma$, $\sigma_{2\text{D}}$, $\tilde \sigma_{2\text{D}}$
corresponding to the luminosities $\tilde L$, $L_{2\text{D}}$, $\tilde
L_{2\text{D}}$ described in Table~\ref{t:survival-notation} and then obtain an
estimation for the cross section taking into account non-electromagnetic
interactions \emph{and} the Pauli form factor as $\sigma = \tilde \sigma \cdot
(\sigma_{2\text{D}} / \tilde \sigma_{2\text{D}})$.

The measured value is the fiducial cross section for the $pp \to pp \mu^+ \mu^-$
reaction with the following constraints:
\label{atlas-constraints}
\begin{itemize}
  \item for $12~\text{GeV} < \sqrt{s} < 30~\text{GeV}$, $p_T > 6$~GeV,
  \item for $30~\text{GeV} < \sqrt{s} < 70~\text{GeV}$, $p_T > 10$~GeV,
  \item $\abs{\eta} < 2.4$,
\end{itemize}
where $\sqrt{s}$ is the invariant mass of the muon pair, $p_T$ and $\eta$ are
the transversal momentum and pseudorapidity of each of the muons.

Calculation of $\tilde \sigma$ is performed with the following code:
\begin{lstlisting}[style=python]
import epa

muon_mass = 105.6583745e-3
def fiducial_xsection(pT_min):
    'Fiducial cross section depending on the constraint on muon momentum'
    return epa.pp_to_ppll(
            13e3,
            mass              = muon_mass,
            pT_min            = pT_min,
            eta_max           = 2.4,
            integration_level = 1
    )

integrate = epa.default_integrator(0)

def calculate(xsection, rs_min, rs_max):
    step = (rs_max - rs_min) / 100
    rs = rs_min
    while rs < rs_max - step:
        print(f'{rs:19.12e} {xsection(rs):19.12e}')
        rs += step
    print(f'{rs_max:19.12e} {xsection(rs_max):19.12e}')
    return integrate(xsection, rs_min, rs_max)

total  = calculate(fiducial_xsection(6),  12, 30)
total += calculate(fiducial_xsection(10), 30, 70)
print(f'# total = {total:19.12e}')
\end{lstlisting}
The function \verb|pp_to_ppll| called at line 6 returns a function computing the
fiducial cross section~\eqref{fiducial-xsection} for $pp$ collisions. The
experiment does not set constraints on photons energies, so
in~\eqref{rapidity-constraints} we have $\hat \omega_{1,\text{min}} = \hat
\omega_{2,\text{min}} = 0$, $\hat \omega_{1,\text{max}} = \hat
\omega_{2,\text{max}} = \infty$ hence $\tilde y = -\infty$, $\tilde Y = \infty$.

Calculation of $\sigma_{2\text{D}}$ can be performed with the same code, with
the function \verb|pp_to_ppll| in line 6 replaced with \verb|pp_to_ppll_b| which
has the same signature.\footnote{
  However, see footnote~\ref{fn:parallelization} on
  page~\pageref{fn:parallelization}. To speed up calculations and avoid roundoff
  errors, we sacrificed accuracy by setting \texttt{relative\_error} to $3^{-n}
  \cdot 10^{-2}$ where $n$ is the integration level. Also, values of the
  differential cross section were stored in an array along with the
  corresponding energies, and the integration was performed over a function
  linearly interpolating between these points.
}

Calculation of $\tilde \sigma_{2\text{D}}$ requires a different function
\verb|fiducial_xsection|:
\begin{lstlisting}[style=python, firstline=4, lastline=16, firstnumber=4]
def fiducial_xsection(pT_min):
    'Fiducial cross section depending on the constraint on muon momentum'
    return epa.xsection_fid(
                epa.photons_to_fermions_pT(muon_mass),
                epa.luminosity_fid(
                    epa.proton_dipole_spectrum_Dirac(6.5e3),
                    integrator = epa.default_integrator(2)
                ),
                mass    = muon_mass,
                pT_min  = pT_min,
                eta_max = 2.4,
                integrator = epa.default_integrator(1)
    )
\end{lstlisting}
The function returned by \verb|photons_to_fermions_pT| calculates a pair of
cross sections~\eqref{photons-to-fermions-parallel-pT},
\eqref{photons-to-fermions-perp-pT}.

\begin{figure}[!tb]
  \includegraphics{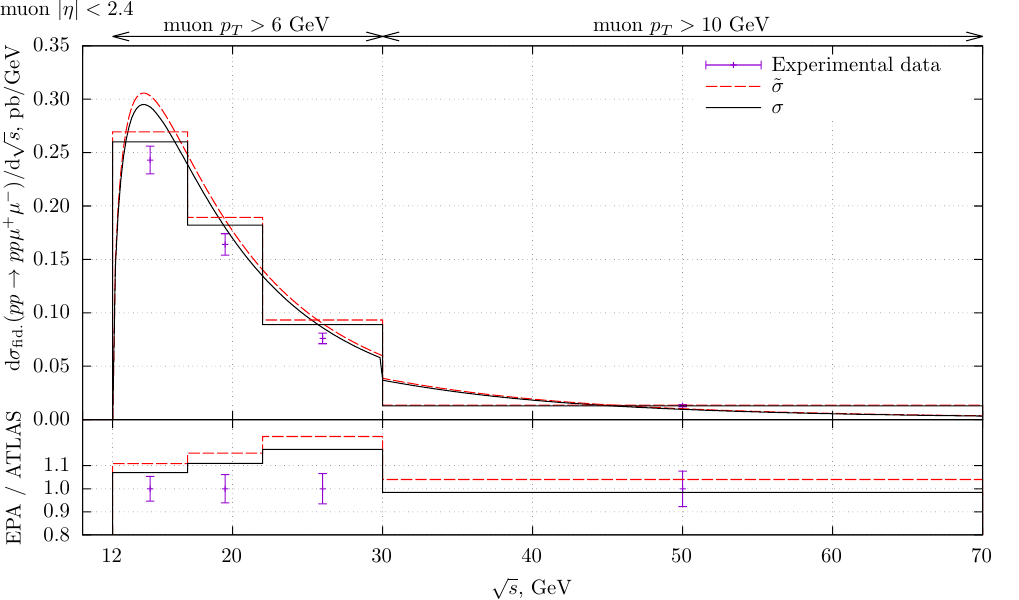}
  \caption{
    \textit{Upper plot:} fiducial cross sections for the reaction $pp \to pp
    \mu^+ \mu^-$ with the collision energy 13~TeV and experimental constraints
    as described in the text at page~\pageref{atlas-constraints} (see also the
    top of the plot). The points are experimental data from
    Ref.~\cite{1708.04053}. $\tilde \sigma$ is the differential cross section
    with non-electromagnetic interactions neglected.  $\sigma$ is that with
    non-electromagnetic interactions taken into account.  The histograms are the
    cross sections integrated over experimental bins.  \textit{Lower plot:}
    ratio of the calculated cross section to the measurement.
  }
  \label{f:atlas}
\end{figure}

Resulting differential cross sections are presented in Fig.~\ref{f:atlas}.
Integrated cross sections are
\begin{itemize}
  \item experimental value: $3.12 \pm 0.07~\text{(stat.)} \pm
  0.10~\text{(syst.)}$~pb~\cite{1708.04053}.
  \item $\tilde \sigma_{2\text{D}} = 3.46$~pb.
  \item $\sigma_{2\text{D}} = 3.31$~pb.
  \item $\tilde \sigma = 3.57$~pb.
  \item $\sigma = 3.44$~pb.
\end{itemize}
Let us compare these results to our previous calculations: 
\begin{itemize}
  \item Ref.~\cite{2106.14842}:
  \begin{itemize}
    \item $\tilde \sigma_{2\text{D}} = 3.39$~pb.
    \item $\sigma_{2\text{D}} = 3.26$~pb.
  \end{itemize}
  The difference here is in the different value of the spectrum parameter
  $\Lambda^2$~\eqref{ff-dipole-lambda}: Ref.~\cite{2106.14842} uses $\Lambda^2 =
  0.61~\text{GeV}^2$, while in this paper we use $\Lambda^2 = 0.66~\text{GeV}^2$
  obtained from the updated value of proton charge radius.
  \item Ref.~\cite{1806.07238}:
  \begin{itemize}
    \item $\tilde \sigma_2 = 3.35$~pb.
  \end{itemize}
  In this case the calculated cross section corresponds to the luminosity
  $\tilde L_2$ in Table~\ref{t:survival-notation}; the calculation used the
  photon spectrum of proton $n_2(\omega)$~\eqref{spectrum-dipole} with even more
  outdated value of $\Lambda^2 = 0.71~\text{GeV}^2$.
\end{itemize}
Ref.~\cite{1708.04053} also presents predictions made with Monte Carlo
simulations:
\begin{itemize}
  \item \texttt{HERWIG}~\cite{0803.0883, 1512.01178}:
  \begin{itemize}
    \item with non-electromagnetic interactions neglected: $3.56 \pm 0.05$~pb.
    \item with non-electromagnetic interactions taken into account with the help
    of corrections from Ref.~\cite{1410.2983}: $3.06 \pm 0.05$~pb.
  \end{itemize}
  \item \texttt{SuperChic2}~\cite{1508.02718}:\footnote{
    See also Table 2 in~\cite{2104.13392}.
  }
  \begin{itemize}
    \item with non-electromagnetic interactions taken into account: $3.45 \pm
    0.05$~pb.
  \end{itemize}
  Note the excellent agreement between \texttt{libepa} and
  \texttt{SuperChic2}.\footnote{
    \texttt{SuperChic2} uses the dipole form factor approximation~\eqref{ff-D},
    \eqref{ff-dipole-sachs} with $\Lambda^2 = 0.71~\text{GeV}^2$.
    \texttt{libepa} cross section $\sigma$ with this $\Lambda$ is $3.50$~pb.
  }
\end{itemize}

\section{Validation}

\label{s:validation}

One important aspect of introducing a new calculation program or library is its
validation by and comparison to the experimental data and other calculation
tools existing in the field. \texttt{libepa} code grew out of calculations
testing theoretical descriptions of the ATLAS measurement of the fiducial cross
section for muon pair production at the LHC which we provide as an example in
Section~\ref{s:example-atlas}. Agreement within 2 standard deviations of the
experimental uncertainty between the measurement results and the library
calculation is already a good validation check. In that calculation we also have
a good agreement in the integrated cross section with \texttt{HERWIG} and
\texttt{SuperChic2}. In this section, let us compare \texttt{libepa} with other
programs in a few other calculations. We follow here Ref.~\cite{2207.03012}
which does a similar comparison for another program, \texttt{gamma-UPC}.

\subsection{Axion production}

\label{s:axion}

Ref.~\cite{2207.03012} uses the program \texttt{gamma-UPC} to calculate
production of axion-like particles in ultraperipheral collision of protons with
the collision energy 14~TeV. Axions are described by the effective Lagrangian
\begin{equation}
    \mathcal{L}
    = \frac12 \partial_\mu a \, \partial^\mu a
    - \frac{m_a^2}{2} a^2
    - \frac{g_{a \gamma}}{4} a F^{\mu \nu} \tilde F_{\mu \nu},
\end{equation}
where $a$ is the axion field, $m_a$ is the axion mass, $g_{a \gamma}$ is the
axion coupling to photons, $F_{\mu \nu}$ is the electromagnetic tensor, $\tilde
F_{\mu \nu}$ is its dual tensor. Cross section for the production of an axion in
a fusion of two photons is
\begin{equation}
    \sigma(\gamma \gamma \to a)
    = \frac{\pi}{16} g_{a \gamma}^2 m_a \delta(\sqrt{s} - m_a).
\end{equation}
Substituting this expression into~\eqref{xsection} and integrating over
$\sqrt{s}$, we get that the production cross section is proportional to the
product of the axion mass and the photon-photon luminosity:
\begin{equation}
  \sigma(pp \to pp a)
  = \frac{\pi}{16} g_{a \gamma}^2 m_a \frac{\mathrm{d} L}{\mathrm{d} \sqrt{s}}.
\end{equation}

Ref.~\cite{2207.03012} uses two equivalent photon spectra, $n_\text{EDFF}$ and
$n_\text{ChFF}$, where ``EDFF'' stands for ``electric dipole form factor'', and
``ChFF'' means ``charge form factor''. For proton-proton collisions, the ChFF
spectrum is the same as our dipole spectrum $n_2$~\eqref{spectrum-dipole},
\eqref{spectrum-dipole-b} with the parameter $\Lambda^2 = 0.71~\text{GeV}^2$.
The EDFF spectrum is defined through the equations\footnote{Cf. Eq.~(11), (12)
in~\cite{2207.03012}; note that \texttt{libepa} and~\cite{2207.03012} use
different normalizations of the equivalent photon spectrum.}
\begin{align}
  n_\text{EDFF}(b, \omega)
  &= \frac{Z^2 \alpha}{\pi^2} \cdot \frac{\omega}{\gamma^2} \left[
         K_1^2 \left( \frac{b \omega}{\gamma} \right)
       + \frac{1}{\gamma^2} K_0^2 \left( \frac{b \omega}{\gamma} \right)
     \right], 
  \label{spectrum-edff-b}
  \\
  n_\text{EDFF}(\omega)
  &= \frac{2 Z^2 \alpha}{\pi \omega}
     \cdot \chi \left[
         K_0(\chi) K_1(\chi)
       - \frac{\chi}{2}
         \left(1 - \frac{1}{\gamma^2} \right)
         \left( K_1^2(\chi) - K_0^2(\chi) \right)
     \right],
     \ \chi = \frac{R_A \omega}{\gamma},
  \label{spectrum-edff}
\end{align}
where $R_A$ is the effective charge radius of the source particle, $R_A =
0.877$~fm\footnote{This value was taken from the \texttt{gamma-UPC} code in
order to compare its results presented in~\cite{2207.03012} to \texttt{libepa}.
The value used by \texttt{libepa} by default is
$0.8414$~fm~\cite{rmp93-025010}.} in the case of proton. This is the spectrum of
a classical charged particle passing by at the ultrarelativistic speed. Its
derivation can be found, e.g., in~\cite[\textsection{}15.5]{jackson}.

Another difference in the calculation described in~\cite{2207.03012} is in the
computation of the parameter $B$ via~\eqref{pp-elastic-slope}. The values of the
parameters used are $B_0 = 9.7511~\text{GeV}^{-2}$, $B_1 = 0.222796~\text{GeV}^{-2}$,
$B_2 = 0.0179103~\text{GeV}^{-2}$.\footnote{Also taken from the
\texttt{gamma-UPC} code for comparison purposes. Ref.~\cite{2207.03012} lists
close but different values.} Although they are quite different from those listed
under~\eqref{pp-elastic-slope}, the value of $B$ is approximately the same for
the collision energies from $\sim 10$~GeV to $\sim 20$~TeV, with a 3\%
difference at 14~TeV.

\begin{figure}[!tb]
  \centering
  \includegraphics{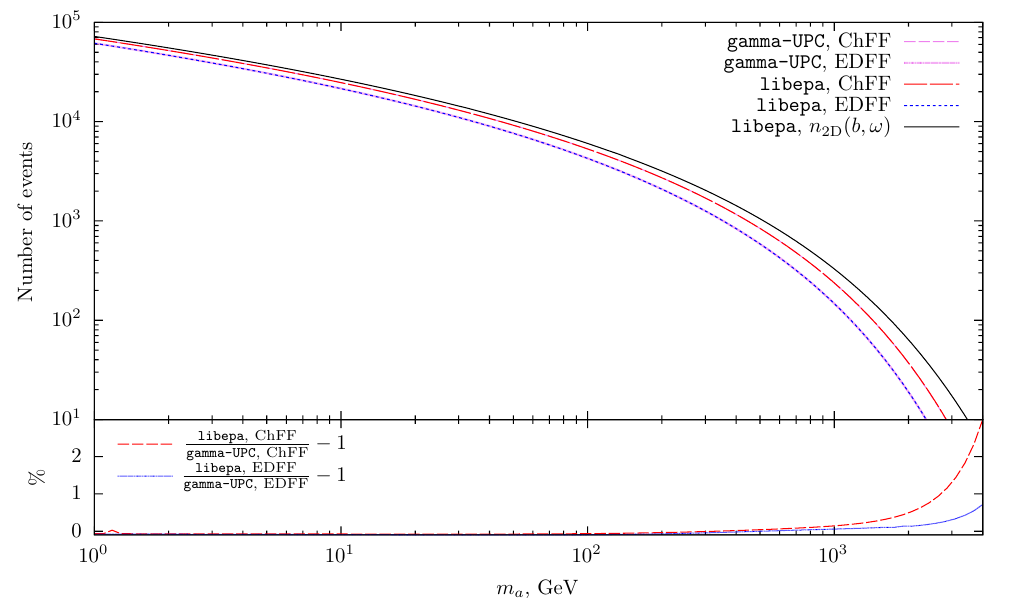}
  \caption{
    Number of events of axion production in $3~\text{ab}^{-1}$ of integrated
    luminosity of ultraperipheral proton-proton collisions with the energy
    14~TeV, assuming axion to photons coupling $g_{a \gamma} =
    0.1~\text{TeV}^{-1}$. \texttt{gamma-UPC} plots are the $pp$ plots presented
    in Fig.~6 of~\cite{2207.03012} but obtained through our own run of
    \texttt{gamma-UPC}. \texttt{libepa} ChFF and EDFF are computed by
    \texttt{libepa} using the same spectra and the same parameter $B$ as in the
    \texttt{gamma-UPC} code. The solid line is computed by \texttt{libepa} using
    the spectrum $n_{2\text{D}}(b, \omega)$~\eqref{spectrum-dirac-b} and the
    default values of the $B$ and $\Lambda$ parameters.
  }
  \label{f:axion}
\end{figure}

We compare the number of events of axion production calculated by
\texttt{gamma-UPC} and \texttt{libepa} in Fig.~\ref{f:axion}. Three calculations
were performed by \texttt{libepa}: two similar to those of \texttt{gamma-UPC},
with the same spectra and the same $B$ parameter, and one with the
\texttt{libepa} default settings. Non-electromagnetic interactions were taken
into account in all cases. \texttt{libepa} and \texttt{gamma-UPC} show excellent
agreement up to the invariant mass of a few TeV.

\subsection{Heavy ions collisions}

\label{s:heavy-ions}

Although \texttt{libepa} was used mostly to calculate cross sections for
proton-proton collisions, its approach is generic enough to be used to deal with
heavy ions as well. However, there is one crucial bit missing, and that is the
probability to avoid non-electromagnetic interactions, $P_{AB}(b)$
in~\eqref{luminosity-b}. The standard approach here is based on the so-called
Glauber formalism~\cite{npb21-135}; see, e.g.,~\cite{nucl-ex/0302016} for a
ready-to-use description. This approach requires calculation of triple integrals
to compute $P_{AB}(b)$, and it drives the level of integration too high to be
practical in terms of numerical accuracy and calculation time. Some of the
future development of \texttt{libepa} certainly lies in this direction.

In the meantime, let us compare calculations of \texttt{libepa} neglecting
non-electromagnetic interactions to other programs that don't do so.
Fig.~8 of~\cite{2207.03012} compares \texttt{gamma-UPC} to \texttt{SuperChic
3.03}~\cite{1810.06567, 2007.12704} and \texttt{StarLight
3.0}~\cite{1607.03838} in electron pair production in ultraperipheral collisions
of lead ions with the energy $5.02$~TeV per nucleon pair. As in the case of
axion production considered above, \cite{2207.03012} provides results for two
spectra, EDFF and ChFF; however in this case the ChFF spectrum is calculated
through~\eqref{spectrum-b-electric}, with the form factor being the Fourier
transform of the Woods-Saxon distribution. The resulting formula is quite
complicated, so we reproduce only the EDFF calculation here,
using~\eqref{spectrum-edff} with $R_A = 7.1$~fm.

Fig.~8 of~\cite{2207.03012} presents two values, one is the cross section
differentiated with respect to the invariant mass of the electron pair, the
other is the cross section differentiated with respect to the pair rapidity.
The phase space is limited by the constraints $p_T > \hat p_T = 2$~GeV,
$\abs{\eta} < \hat \eta = 2.4$, $\sqrt{s} > 5$~GeV.  Calculation of the former
cross section is a straightforward application of~\eqref{fiducial-xsection}. The
latter requires swapping of the integrations with respect to $p_T$ and $y$.
Beginning with~\eqref{eta-y} and treating the constraint $-\hat \eta < \eta <
\hat \eta$ as a constraint on $p_T$, we get
\begin{multline}
  \frac{\mathrm{d} \sigma_\text{fid.}(AB \to AB \chi^+ \chi^-)}
       {\mathrm{d} y \, \mathrm{d} \sqrt{s}}
  = \int\limits_{\max(\hat p_T, P_T(\abs{y}))}
               ^{\frac{\sqrt{s}}{2} \sqrt{1 - \frac{4 m_\chi^2}{s}}}
      \mathrm{d} p_T
      \frac{\mathrm{d} \sigma(\gamma \gamma \to \chi^+ \chi^-)}{\mathrm{d} p_T}
      \frac{\mathrm{d} L_{AB}}
           {\mathrm{d} y \, \mathrm{d} \sqrt{s}}
  \\
  \times \theta \left(
    \sqrt{1 - \frac{4 m_\chi^2}{s}} - \frac{\sinh \abs{y}}{\sinh \hat \eta}
  \right),
  \label{xsection/rapidity}
\end{multline}
where $\theta(x)$ is the Heaviside step function, and
\begin{equation}
  \frac{\mathrm{d} L_{AB}}{\mathrm{d} y \, \mathrm{d} \sqrt{s}}
  = \frac{\sqrt{s}}{2} \,
    n_A \left( \frac{\sqrt{s}}{2} \mathrm{e}^y \right) \,
    n_B \left( \frac{\sqrt{s}}{2} \mathrm{e}^{-y} \right)
\end{equation}
is the photon-photon luminosity differentiated with respect to both the
invariant mass and the rapidity of the produced system, available in
\texttt{libepa} by means of the \texttt{luminosity\_y} function.

\begin{figure}[!tb]
  \centering
  \includegraphics{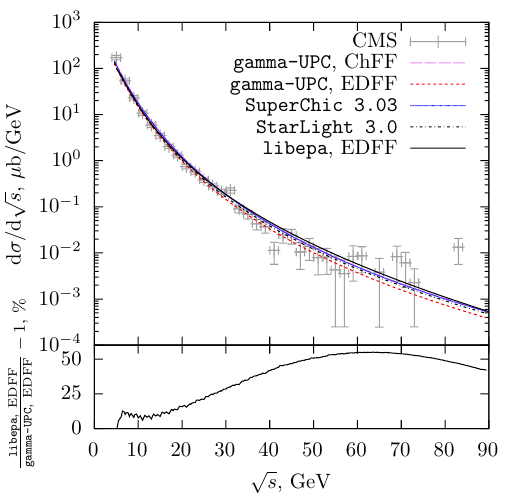}\hspace*{-10pt}\includegraphics{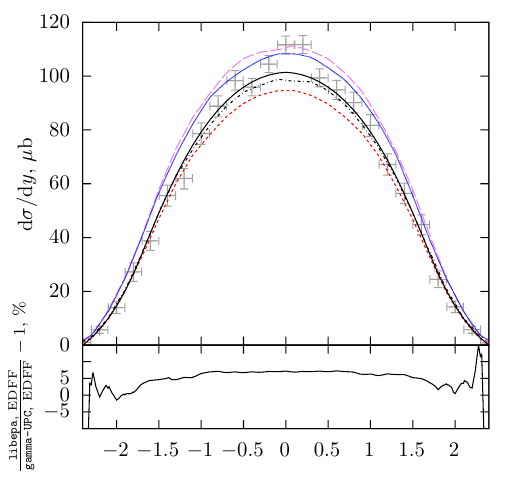}
  \caption{
    Fiducial cross section for the production of a pair of electrons in
    ultraperipheral collisions of \nuc{Pb}{208} nuclei with the collision energy
    5.02~TeV/(nucleon pair). \textit{Left:} cross section differentiated with
    respect to the invariant mass of the electron pair. \textit{Right:} cross
    section differentiated with respect to the rapidity of the electron pair.
    The experimental points and the \texttt{gamma-UPC}, \texttt{SuperChic} and
    \texttt{StarLight} lines were taken from Fig.~8 of~\cite{2207.03012}. The
    jitter in the ratio plots is due to poor digitization accuracy.
  }
  \label{f:electrons}
\end{figure}

Results of the calculations are shown in Fig.~\ref{f:electrons} along with the
points and lines copied from Fig.~8 of~\cite{2207.03012} with the help
of~\cite{web-plot-digitizer}. Note that the CMS data~\cite{1810.04602} is scaled
in~\cite{2207.03012} to fit the \texttt{StarLight} calculations.  The integrated
cross section computed by \texttt{libepa} is 289~$\mu$b; calculation results
listed in Table~XII of~\cite{2207.03012} are: \texttt{gamma-UPC} EDFF~---
272~$\mu$b, ChFF~--- 326~$\mu$b, \texttt{StarLight}~--- 285~$\mu$b,
\texttt{SuperChic}~--- 318~$\mu$b. Since \texttt{libepa} does not take into
account non-electromagnetic interactions while all the other calculations do so,
\texttt{libepa} result is expected to be larger than \texttt{gamma-UPC} EDFF.
Most important part of the integration domain is at $5~\text{GeV} < \sqrt{s}
\lesssim 10~\text{GeV}$; according to Fig.~4 from~\cite{2207.03012},
non-electromagnetic interactions reduce the cross section by $\sim 10$\% in this
region. The fact that \texttt{libepa} result fits so nicely the scaled
experimental points appears to be a consequence of EDFF
spectrum~\eqref{spectrum-edff} being less than the equivalent photon spectrum
used in other calculations.

\section{Conclusions}

The design of \texttt{libepa} is quite different from other programs used to
calculate cross sections of ultraperipheral collisions~\cite{1102.2531,
1607.03838, 2007.12704, 2207.03012}. The key differences are:
\begin{enumerate}
  \item \texttt{libepa} is a library rather than a standalone program.
  \item \texttt{libepa} relies on deterministic one-dimensional integration
  rather than the Monte Carlo approach.
  \item \texttt{libepa} is designed in the functional programming paradigm.
\end{enumerate}
Being a library, \texttt{libepa} provides a set of tools for the user to create
their own computation rather than a set of pre-programmed computations with
variable numerical parameters. The functional programming approach allowed for
the interface when the user can replace part of a common computation with their
own function, e.g., by changing the spectrum of a colliding particle, tweak the
integration algorithm, or build a computation for a function not explicitely
supported by the library. An example of the latter may be the computation of the
rapidity distribution shown in the right plot in Fig.~\ref{f:electrons}~--- at
present there is no function in \texttt{libepa} that would readily provide cross
section~\eqref{xsection/rapidity}, but it can be easily constructed using other
\texttt{libepa} functions, see \texttt{electrons.cpp} in the supplementary
materials. At the same time common computations are kept simple, and cross
sections for proton-proton collisions discussed in Section~\ref{s:upc-xsection}
can be obtained by a single call to \texttt{libepa}.

Functional programming paradigm has granted \texttt{libepa} two more interesting
features that would be difficult to implement without. One is that there is a
precomputation stage while building the computation function that allows for
optimizations relevant to the problem at hand, such as calculation of values
that remain constant during the main execution, allocation of memory for
integration algorithms or elimination of redundant code jumping across the
boundary of the foreign function interface. These optimizations are transparent
to the user of the library. The other is that combined with CFFI bindings to a
language that features a read-evaluate-print loop (REPL), it gives the user a
powerful calculator that can quickly evaluate various values of interest to the
research at hand. The library initially was being developed in Common Lisp, and
although Common Lisp bindings haven't made it into \texttt{libepa} release (they
might in the future), those to Python coupled with Python REPLs or notebooks
should provide a similar experience.

The fact that \texttt{libepa} uses deterministic integration rather than Monte
Carlo may be an advantage or a disadvantage depending on the problem at hand and
the approach to solve it. An explicit representation of the computation function
in terms of mathematical expressions possibly involving recurring
one-dimensional integrals over well-defined domains is required. This is
particularly problematic for calculations of fiducial cross sections of
processes with more than two particles in the final state. However, once such an
expression is obtained, \texttt{libepa} may be an excellent tool to test and use
it, since all that is needed is to program the mathematical expressions,
preferably keeping an eye open for optimizations such as avoiding redundant
calculations.

One can expect that deterministic integration may take less time than the Monte
Carlo approach. This is probably true as long as the integration level isn't too
high, but direct comparison to other programs is complicated since they often
compute multiple quantities at once, some of them cache intermediate results to
improve calculation time when less important variables change, and we would also
need to study carefully the calculation uncertainty, to which the calculation
time is very sensitive in the Monte Carlo approach. That would require a
dedicated research that is not at present of interest to the authors of this
paper.

\texttt{libepa} approaches and code were developed while the authors were
working on papers~\cite{1806.07238, 1906.08568, 2012.01599, 2106.14842}. It was
applied to semi-inclusive processes (where only one of the colliding particles
remains intact, and the other disintegrates) in papers~\cite{2207.07157,
2308.01169}.

The authors were supported by the Russian Science Foundation grant
No~19-12-00123-$\Pi$.

%\section{Conclusions}
%
%\texttt{libepa} approaches and code were developed while the authors were
%working on papers~\cite{1806.07238, 1906.08568, 2012.01599, 2106.14842}. It was
%applied to semi-inclusive processes (where only one of the colliding particles
%remains intact, and the other disintegrates) in papers~\cite{2207.07157,
%2308.01169}. Functional programming approach to calculations used in
%\texttt{libepa} proved to be flexible and allowing for quick tests of various
%assumptions. Initially the library was being developed in Common Lisp
%programming language, and when coupled with the Common Lisp read-evaluate-print
%loop (REPL) it gave the user a powerful calculator that could quickly evaluate
%various values of interest to the research at hand. Although Common Lisp
%bindings haven't made it into \texttt{libepa} release (they might in the
%future), those to Python coupled with Python REPLs or notebooks should provide
%a similar experience.
%
%Another benefit of the  functional programming paradigm is that there is a
%precomputation stage while building the computation function that allows for
%optimizations relevant to the problem at hand, such as calculation of values
%that remain constant during the main execution, allocation of memory for
%integration algorithms or elimination of redundant code jumping across
%the boundary of the foreign function interface. These optimizations are
%transparent to the user of the library.
%
%The authors were supported by the Russian Science Foundation grant
%No~19-12-00123-$\Pi$.

\newcommand{\arxiv}[1]{arXiv:\nolinebreak[3]\href{http://arxiv.org/abs/#1}{#1}}

\end{document}